\documentclass[12pt]{article}
\usepackage{epsfig,here,subeqn}
\usepackage{graphicx}

\newcommand{\be}{\begin{equation}}
\newcommand{\ee}{\end{equation}}
\newcommand{\bea}{\begin{eqnarray}}
\newcommand{\eea}{\end{eqnarray}}
\newcommand{\ba}{\begin{array}}
\newcommand{\ea}{\end{array}}
\newcommand{\no}{\nonumber}
\newcommand{\aki}{\hspace*{12pt}}
\newcommand{\dsl}{\hspace{-5.5pt}/}
\newcommand{\cdt}{\!\cdot\!}

\begin{document}
\begin{titlepage}
\title{Tau Polarization in Tau-Neutrino Nucleon Scattering}
\author{
{Kaoru Hagiwara}\\
{\normalsize \sl Theory Group, KEK, Tsukuba 305-0801, JAPAN}\\
{}\\
{Kentarou Mawatari\footnote{E-mail address: 
                   \texttt{mawatari@radix.h.kobe-u.ac.jp}}}\\
{\normalsize \sl Graduate School of Science and Technology,
Kobe University,}\\
{\normalsize \sl Nada, Kobe 657-8501, JAPAN}\\ 
{}\\
{Hiroshi Yokoya\footnote{E-mail address: 
           \texttt{yokoya@theo.phys.sci.hiroshima-u.ac.jp}}}\\
{\normalsize \sl Department of Physics, Hiroshima University,}\\
{\normalsize \sl Higashi-Hiroshima 739-8526, JAPAN}\\
{\normalsize \sl and Radiation Laboratory, RIKEN,
 Wako 351-0198, JAPAN}
}
\date{}
\maketitle

\begin{abstract}
{\normalsize 
We investigate the spin polarization of $\tau^{\pm}$ leptons
produced in $\nu_{\tau}$ and $\bar{\nu}_{\tau}$ nucleon scattering 
via charged currents. Quasi-elastic scattering, 
$\Delta$ resonance production and deep inelastic scattering 
processes are studied.  
The polarization information is essential for measuring the $\tau^{\pm}$ 
appearance rate in long baseline neutrino oscillation experiments, 
because the decay particle distributions depend crucially on the 
$\tau^{\pm}$ spin. In this article, we calculate the spin density 
matrix of each process and estimate the spin polarization vector in 
medium and high neutrino energy interactions. 
We find that the produced $\tau^{\pm}$'s have 
high degree of polarization, and their spin direction depends 
non-trivially on the energy and the scattering angle of 
$\tau^{\pm}$ in the laboratory frame.}
\end{abstract}

\begin{picture}(5,2)(20,-650)
\put(350,-100){KEK-TH-887}
\put(350,-115){KOBE-FHD-03-01}
\put(350,-130){HUPD-0303}
\put(350,-145){hep-ph/0305324}
\put(350,-160){May 2003}
\end{picture}

\thispagestyle{empty}
\end{titlepage}
\setcounter{page}{1}

\section{Introduction}
\hspace*{\parindent}
Recent studies from neutrino oscillation experiments
are revealing the amazing nature of the neutrino sector, 
with their non-zero masses and large flavor mixings.
Especially, reports from Super-Kamiokande (SK)
collaboration\cite{sk} strongly suggest that nearly maximal 
oscillation from $\nu_{\mu}$ into $\nu_{\tau}$ is occurring
in the atmospheric neutrino flux. 
To demonstrate this oscillation, it is important 
to detect $\nu_{\tau}$ appearance in oscillation experiments.
Several long-baseline neutrino oscillation 
experiments, such as ICARUS\cite{icarus}, 
MINOS\cite{minos}, OPERA\cite{opera} are proposed, 
and they are expected to detect the $\tau$ appearance by 
charged current (CC) reactions off a nucleon target
\bea
\nu_{\tau}(\bar{\nu}_{\tau}) + N \to \tau^{-}(\tau^{+}) + X
\eea
with $N=p, n$. Because $\tau$ production by a nucleon target 
has a threshold for neutrino energy at $E_{\nu}\approx 3.5$GeV,
these experiments should provide high energy neutrino flux. 
It has also been pointed out by Hall and Murayama\cite{hm}, 
that SK may be able to detect the $\tau$ appearance events 
with more than several years of running.\\

The produced $\tau$ decays into several particles, 
always including a neutrino ($\nu_{\tau}$). 
Therefore the $\tau$ appearance signal should be obtained 
from decay particle distributions. 
Because the $\tau$ decay distributions depend
significantly on it's spin polarization\cite{tauola}, 
the polarization information is essential 
for us to identify the $\tau$ production signal.\\

$\tau$ polarization should also be studied in order to
estimate background events for the 
$\nu_{\mu}\to\nu_{e}$ appearance reaction, 
which will be searched for in neutrino oscillation experiments, 
such as those using high intensity neutrino beams from 
J-PARC\cite{J-PARC}.
Because the oscillation amplitude of $\nu_{\mu} \to \nu_{\tau}$ 
is larger than that of $\nu_{\mu} \to \nu_{e}$\cite{chooz}, 
and because the branching ratio of $\tau^{\mp}\to e^{\mp}+X$ is 
relatively large, the $e$ production via the 
$\nu_{\mu} \to \nu_{\tau} \to \tau \to e$ 
chain can be significant\cite{aoki}.
Since the $e$ energy and angular distribution depends 
crucially on the $\tau$ polarization, it's information is 
necessary to estimate the background.\\

So far, several authors have calculated the $\tau$ 
production cross section for nucleon targets\cite{hm,py,kretzer}, 
but to our knowledge, no estimation of the polarization of 
produced $\tau$'s is available. 
In this paper, we study the spin polarization 
of $\tau$ produced by $\nu_{\tau}$ scattering off a nucleon 
target. We consider the quasi-elastic scattering (QE), 
$\Delta$ resonance production (RES), and deep inelastic 
scattering (DIS) processes, which are known 
to give dominant contributions in the medium 
and high neutrino energy region\cite{py}. 
The spin polarization vector is obtained 
from the spin density matrix which is calculated
for each process.\\

The article is organized as follows. 
We give the general kinematics of $\tau$ production 
 in neutrino-nucleon interaction and the relation 
between the spin density matrix 
and the spin polarization vector in section \ref{seckine}. 
Then we present the details of the spin density matrix calculation 
of QE, RES, and DIS processes, in sections \ref{secqe}, 
\ref{secres}, and \ref{secdis}, respectively.  
In section \ref{result}, the differential cross section and the  
spin polarization vector of produced 
$\tau^{\pm}$ are estimated for medium and high neutrino energies. 
Section \ref{discussion} gives discussions and our conclusions.

\section{Kinematics and Formalism}\label{seckine}
\hspace*{\parindent}
In this section, we show the physical regions of 
kinematical variables and give the relation of the $\tau$ 
spin polarization vector and the spin density matrix 
of the charged current (CC) $\tau$ production process. 
Firstly, we define the four-momenta of incoming neutrino ($k$), 
target nucleon ($p$) and produced $\tau$ lepton ($k'$) in the 
laboratory frame
\bea
k^{\mu}\, \!\!\!&=&\!\! (E_{\nu},0,0,E_{\nu}),\\
p^{\mu}\, \!\!\!&=&\!\!  (M,0,0,0),\\
k'^{\mu}  \!\!\!&=&\!\!  (E_{\tau},p_{\tau}\sin\theta,0,p_{\tau}\cos\theta).
\eea
Here, $E_{\nu}$ and $E_{\tau}$ are the incoming neutrino and 
outgoing $\tau$ 
 energies, respectively, in the laboratory frame, 
$M$ is the nucleon mass, and $p_{\tau}=\sqrt{E_{\tau}^{2}-m_{\tau}^{2}}$ 
with the $\tau$ lepton mass $m_\tau =1.78$GeV. 
We also define some Lorentz invariant variables
\bea
&&Q^{2} = -q^{2}, \aki  q^{\mu}= k^{\mu}-k'^{\mu},\\
&&W^{2} =(p+q)^{2}.
\eea
$Q^{2}$ is the magnitude of the momentum transfer and $W$ is the hadronic 
invariant mass. The physical regions of these variables 
are given by 
\bea
&&M\leq W \leq\sqrt{s}-m_{\tau},\label{wregion}
\eea
and
\bea
&&Q^{2}_{-}(W)\leq Q^{2}\leq Q^{2}_{+}(W),\label{qregion}
\eea
where $s=(k+p)^{2}$ and 
\bea
&&Q^{2}_{\pm}(W)=\frac{s-M^{2}}{2}(1\pm\bar{\beta})
-\frac{1}{2}\left[W^{2}+m^{2}_{\tau}
-\frac{M^{2}}{s}\left(W^{2}-m^{2}_{\tau}\right)\right]
\eea
with $\bar{\beta}=\lambda^{\frac{1}{2}}\left(1,m^{2}_{\tau}/s,W^{2}/s\right)$  
and $\lambda(a,b,c)=a^{2}+b^{2}+c^{2}-2(ab+bc+ca)$.\\

The scaling variables are defined as usual:
\bea
&&x = \frac{Q^{2}}{2\,p\cdt q}=\frac{Q^{2}}{W^{2}+Q^{2}-M^{2}},\\ 
&&y = \frac{p\cdt q}{p\cdt k}=\frac{W^{2}+Q^{2}-M^{2}}{s-M^{2}}
=1-\frac{E_{\tau}}{E_{\nu}}.
\eea
Here, $x$ is the Bjorken variable and $y$ is the inelasticy.  
The physical regions for $x$ and $y$ are obtained by 
Albright and Jarlskog\cite{kretzer,albright}:
\bea
\frac{m_{\tau}^{2}}{2M(E_{\nu}-m_{\tau})}\le x \le 1\label{aj1}
\eea
and
\bea
A-B \le y \le A+B,\label{aj2}
\eea
where
\bea
&&A=\frac{1}{2}\left(1-\frac{m_{\tau}^{2}}{2ME_{\nu}x}
-\frac{m_{\tau}^{2}}{2E_{\nu}^{2}}\right)
\bigg{/}\left(1+\frac{xM}{2E_{\nu}}\right),\label{aj3}\\
&&B=\frac{1}{2}\left[\left(1-\frac{m_{\tau}^{2}}{2ME_{\nu}x}\right)^{2}
-\frac{m_{\tau}^{2}}{E_{\nu}^{2}}\,\right]^{\frac{1}{2}}
\Bigg{/}\left(1+\frac{xM}{2E_{\nu}}\right). \label{aj4}
\eea
The above regions agree with those determined by 
Eq.(\ref{wregion}) and Eq.(\ref{qregion}).\\
 
We label the relevant subprocesses by using the hadronic 
invariant mass $W$ and the momentum transfer $Q^{2}$. 
We label QE (quasi-elastic scattering) when $W=M$, 
RES (resonance production) when $M+m_{\pi}<W<W_{\rm cut}$, 
and IS (inelastic scattering) when  $W_{\rm cut}<W<\sqrt{s}-m_{\tau}$. 
$W_{\rm cut}$ is an artificial boundary between RES and IS processes, 
to avoid double counting. 
The $W_{\rm cut}$ value is taken in the region 1.4GeV$\sim$1.6GeV.
Within the IS region, the region where $Q^{2}\ge 1{\rm GeV}^{2}$
may be labeled as DIS, where the use of the parton model can be 
justified.\\

\begin{figure}[H]
\begin{center}
\epsfig{figure=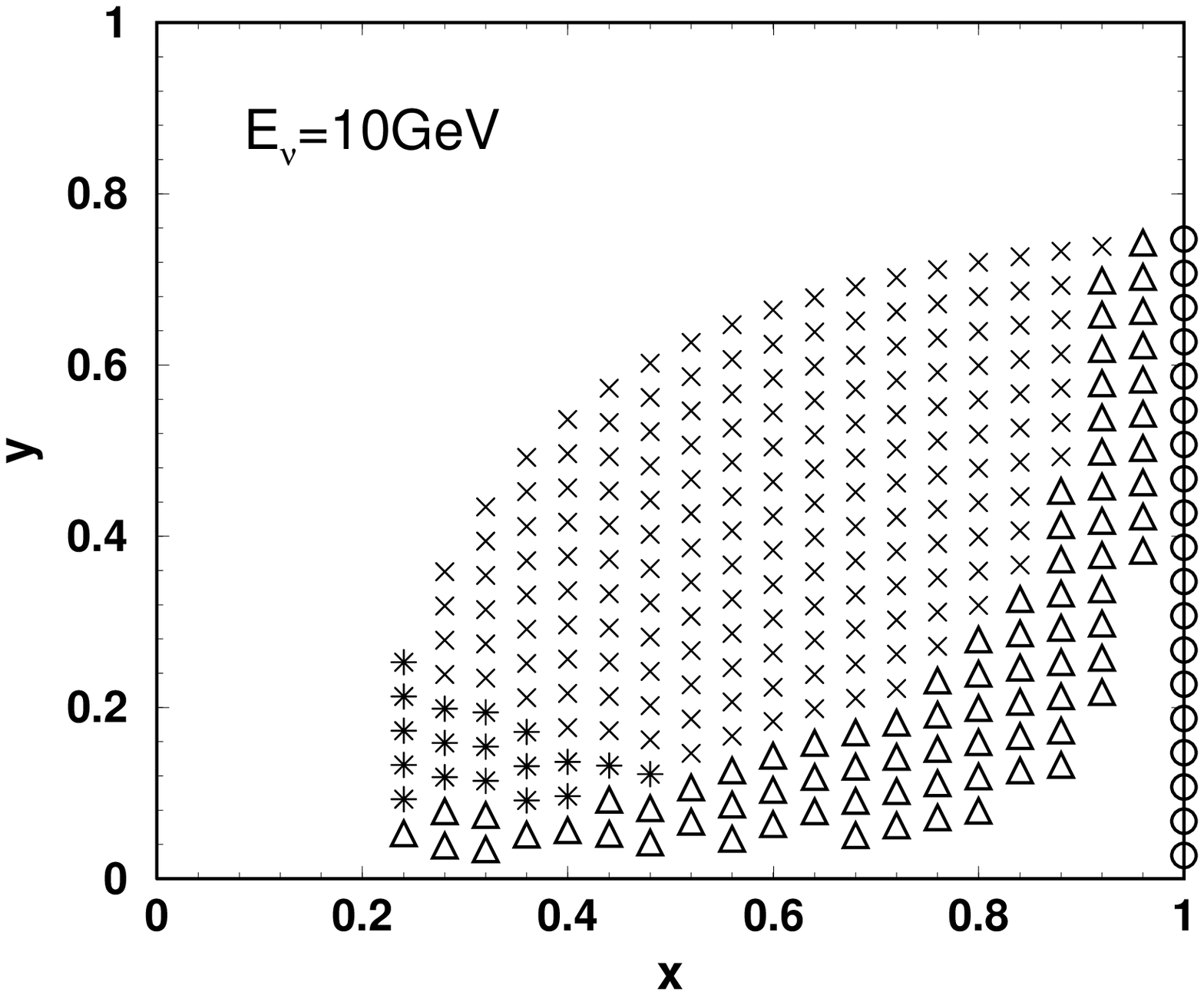,width=6.5cm}
\aki\epsfig{figure=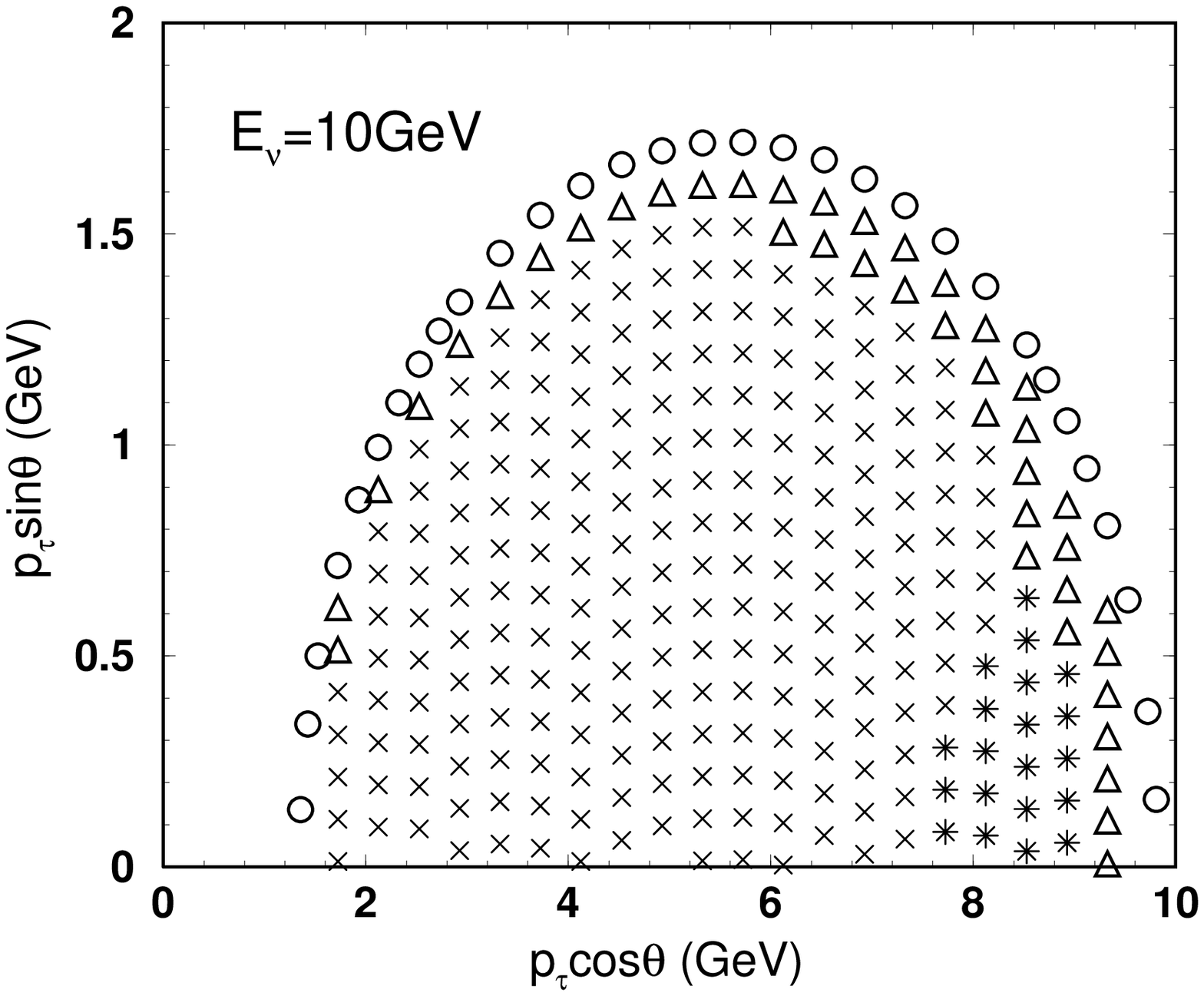,width=6.5cm}
\caption{Physical region at $E_{\nu}=10$ GeV in the $x$-$y$ 
plane (left), and in the $p_{\tau}\cos\theta$-$p_{\tau}\sin\theta$ 
plane (right). Open circles denote QE (quasi-elastic scattering), 
open triangles denote RES ($\Delta$ resonance production),  
and the cross symbols give the DIS (deep inelastic scattering) 
region with $Q^{2}\ge 1 \rm GeV^{2}$. 
The region marked by the star symbol ($\ast$) gives inelastic 
scattering (IS), at $W\ge 1.4$GeV and $Q^{2}<1{\rm GeV}^{2}$.
}\label{kinema10}
\end{center}
\end{figure}

Fig.\ref{kinema10} shows the kinematical regions of 
each QE, RES and IS process on the 
$x$-$y$ plane (left) and the $p_{\tau}\cos\theta$-$p_{\tau}\sin\theta$ 
plane (right) at $E_{\nu}= 10$ GeV. 
The QE region is shown by open circles, the RES region by open triangles,
and the DIS region is shown by the cross symbols. The region shown by
the star symbol ($\ast$) gives the IS process 
at low $Q^{2}$ ($Q^{2}<1{\rm GeV}^{2}$). In this region the parton model 
is not reliable and we must use the experimental data to reduce errors.
In this report, however, we use the parton model throughout the IS region. 
Studies on uncertainties in this region will be reported elsewhere. \\

Produced $\tau$ will be partially polarized. 
We define the spin polarization vector, parameterized as
\bea
\vec{s}=(s_{x},s_{y},s_{z})=
\frac{P}{2}\left(\sin\theta_{P}\cos\varphi_{P},
\sin\theta_{P}\sin\varphi_{P},\cos\theta_{P}\right)
\label{spinvector}\eea
in the $\tau$ rest frame in which the z-axis is 
taken along it's momentum direction in the laboratory frame.
In Eq.(\ref{spinvector}), $\theta_{P}$ and $\varphi_{P}$ are 
the polar and azimuthal angle of the spin vector 
in the $\tau$ rest frame, respectively, 
and $P$ denotes the degree of polarization.
$P=1$ gives the fully polarized $\tau$, and
 $P=0$ gives unpolarized $\tau$. 
The azimuthal angle is measured from the scattering plane where
$\varphi_{P}={\pi}/{2}$ is along the 
$\vec{p}_{\nu}\times\vec{p}_{\tau}$ direction 
in the laboratory frame. The degree of polarization ($P$) and 
the spin directions ($\theta_{P}, \varphi_{P}$)
are functions of $E_{\tau}$ and $\cos\theta$.
This spin polarization vector is related with 
 the spin density matrix $R_{\lambda\lambda'}$, 
by the following relation:
\bea
\frac{dR_{\lambda\lambda'}}{dE_{\tau}\,d\cos\theta}
=\frac{1}{2}
\left(\ba{cc}1+{P}\cos\theta_{P}\ & 
{P}\sin\theta_{P}\,e^{i\varphi_{P}}\\
{P}\sin\theta_{P}\,e^{-i\varphi_{P}}\ & 
1-{P}\cos\theta_{P}\ea \right)\cdt
\frac{d\sigma_{\rm sum}}{dE_{\tau}\,d\cos\theta}.\label{density}
\eea
The density matrix is calculated as
$R_{\lambda\lambda'}\propto \sum M_{\lambda}M^{*}_{\lambda'}$, 
where $M_{\lambda}$ is the helicity amplitude 
with the $\tau$ helicity $\lambda/2$  
defined in the laboratory frame,
and $d\sigma_{\rm sum}= dR_{++}+dR_{--}$ 
is the usual spin summed cross section.
The summation symbol implies the summation over final states, 
and the spins of the target and final-state particles.\\

The spin density matrix of $\tau$ production 
is obtained by using the leptonic and hadronic tensor as 
\be
\frac{dR_{\lambda\lambda'}}{dE_{\tau}\, d\cos\theta}
= \frac{G_{F}^{2}\kappa^{2}}{4\pi}
\frac{p_{\tau}}{ME_{\nu}}\,
L^{\mu\nu}_{\lambda\lambda'}W_{\mu\nu},\label{density2}
\ee
where $G_{F}$ is Fermi constant and 
$\kappa=M_{W}^{2}/(Q^{2}+M_{W}^{2})$ is the propagator factor 
with the $W$-boson mass $M_{W}=80.4$GeV. 
For $\tau^{-}$ production, the leptonic tensor 
$L^{\mu\nu}_{\lambda\lambda'}$ is expressed as 
\bea
L_{\lambda\lambda'}^{\mu\nu}=j_{\lambda}^{\mu}\,j^{*\nu}_{\lambda'},
\label{lepton}
\eea
where the leptonic weak current $j_{\lambda}^{\mu}$ is 
\bea
j_{\lambda}^{\mu}\!\!\!&=&\!\!\! 
\bar{u}_{\tau}(k',\lambda)\,\gamma^{\mu}
\frac{1-\gamma_{5}}{2}\, u_{\nu}(k)\no\\ 
\!\!\!&=&\!\!\!\left\{\ba{ll}
\sqrt{2E_{\nu}(E_{\tau}-p_{\tau})}\,
\left(\sin\!\frac{\theta}{2},-\cos\!\frac{\theta}{2},
i\cos\!\frac{\theta}{2},\sin\!\frac{\theta}{2}\right)
&\ \ (\lambda=+)\\
\sqrt{2E_{\nu}(E_{\tau}+p_{\tau})}\,
\left(\cos\!\frac{\theta}{2},\sin\!\frac{\theta}{2},
-i\sin\!\frac{\theta}{2},\cos\!\frac{\theta}{2}\right)
&\ \ (\lambda=-)
\ea\right.\label{lepvec}
\eea
in the laboratory frame. 
For $\tau^{+}$ production, 
we must replace the leptonic tensor 
$L^{\mu\nu}_{\lambda\lambda'}$ into 
$\overline{L}^{\mu\nu}_{\lambda\lambda'}$ defined as
\bea
\overline{L}^{\mu\nu}_{\lambda\lambda'}=
\bar{j}^{\mu}_{\lambda}\,\bar{j}^{*\,\nu}_{\lambda'},
\label{antilep}\eea
where $\bar{j}_{\lambda}^{\mu}$ is 
\bea
\bar{j}_{\lambda}^{\mu}\!\!\!&=&\!\!\! \bar{v}_{\nu}(k)\,\gamma^{\mu}
\frac{1-\gamma_{5}}{2}\,v_{\tau}(k',\lambda)\no\\
\!\!\!&=&\!\!\! \left\{ \ba{ll} \sqrt{2E_{\nu}(E_{\tau}+p_{\tau})}
\,\left(\cos\!\frac{\theta}{2},\sin\!\frac{\theta}{2},
i\sin\!\frac{\theta}{2},\cos\!\frac{\theta}{2}\right) &\ \ (\lambda=+)\\
\sqrt{2E_{\nu}(E_{\tau}-p_{\tau})}
\,\left(-\sin\!\frac{\theta}{2},\cos\!\frac{\theta}{2},
i\cos\!\frac{\theta}{2},-\sin\!\frac{\theta}{2}\right) &\ \ (\lambda=-),\\
\ea \right.
\eea
which is related with $j_{\lambda}^{\mu}$ by $\bar{j}_{\lambda}^{\mu}=
\lambda\,j_{-\lambda}^{*\mu}$, in the phase convention of Ref.\cite{hz}.
In the following sections, we will abbreviate the overline of 
the leptonic tensor and currents for $\tau^{+}$ production process.\\

The hadronic tensor is expressed in general as
\bea
W_{\mu\nu}(p,q)=-g_{\mu\nu}W_{1}(p\cdt q,Q^{2})
+\frac{p_{\mu}p_{\nu}}{M^{2}}\,W_{2}(p\cdt q,Q^{2})
-i\epsilon_{\mu\nu\alpha\beta}\frac{p^{\alpha}q^{\beta}}
{2M^{2}}\,W_{3}(p\cdt q,Q^{2})\no\\
+\frac{q_{\mu}q_{\nu}}{M^{2}}\,W_{4}(p\cdt q,Q^{2})
+\frac{p_{\mu}q_{\nu}+q_{\mu}p_{\nu}}{2M^{2}}\,W_{5}(p\cdt q,Q^{2}),
\label{hadten}
\eea
where the totally anti-symmetric tensor $\epsilon_{\mu\nu\alpha\beta}$ is 
defined as $\epsilon_{0123}=1$, 
and the structure functions $W_{i=1,\ldots,5}(p\cdot q,Q^{2})$ 
can be estimated for each subprocess. 
Since $q\cdot j_{\lambda}$ is proportional to $m_{\tau}$, 
the structure functions $W_{4}$ and $W_{5}$ appear only 
in the heavy lepton production case\cite{albright}.\\
  
Inserting these equations into Eq.(\ref{density2}) 
and Eq.(\ref{density}), we find 
\bea
\frac{d\sigma_{\rm sum}}{dE_{\tau}\,d\cos\theta}\!\!\!&=&\!\!\!
\frac{G_{F}^{2}\kappa^{2}}{2\pi}
\frac{p_{\tau}}{M}\,\bigg\{
\Big(2W_{1}+\frac{m_{\tau}^{2}}{M^{2}}\,W_{4}\Big)
\left(E_{\tau}-p_{\tau}\cos\theta\right)
+W_{2}\left(E_{\tau}+p_{\tau}\cos\theta\right)
\no\\&&\hspace*{50pt}
\pm\frac{W_{3}}{M}\,\Big(E_{\nu}E_{\tau}+p_{\tau}^{2}
-(E_{\nu}+E_{\tau})p_{\tau}\cos\theta\Big)
-\frac{m_{\tau}^{2}}{M}\,W_{5}\bigg\}\no\\
\!\!\!&\equiv&\!\!\!\frac{G_{F}^{2}\kappa^{2}}{2\pi}
\frac{p_{\tau}}{M}\;F, \label{cross} 
\eea
and the spin polarization vector takes  
\begin{subequations}
\bea
s_{x} \!\!\!&=&\!\!\! \mp\,\frac{m_{\tau}\sin\theta}{2}
\bigg(2W_{1}-W_{2}\pm\frac{E_{\nu}}{M}\,W_{3}
-\frac{m_{\tau}^{2}}{M^{2}}\,W_{4}+\frac{E_{\tau}}{M}\,W_{5}\bigg)
\bigg{/}F,\\
s_{y}\!\!\!&=&\!\!\!0,\\
s_{z}\!\!\!&=&\!\!\! \mp\,\frac{1}{2}\bigg\{
\Big(2W_{1}-\frac{m_{\tau}^{2}}{M^{2}}\,W_{4}\Big)
\left(p_{\tau}-E_{\tau}\cos\theta\right)
+W_{2}\left(p_{\tau}+E_{\tau}\cos\theta\right)
\no\\&&\hspace*{30pt}
\pm\frac{W_{3}}{M}\,\Big((E_{\nu}+E_{\tau})p_{\tau}
-(E_{\nu}E_{\tau}+p_{\tau}^{2})\cos\theta\Big)
-\frac{m_{\tau}^{2}}{M}\,W_{5}\cos\theta\bigg\}\bigg{/}F,
\eea \label{sss}
\end{subequations}
 for $\tau^{\mp}$ productions. The degree of polarization is given by
\bea
P=2\sqrt{s_{x}^{2}+s_{y}^{2}+s_{z}^{2}}. \label{degreeP}
\eea
The above results Eq.(\ref{cross})-(\ref{degreeP}) agree with 
Ref.\cite{albright}.
From the above equations we find (i) $\varphi_{P}$ takes either  
0 or $\pi$ for any $\tau$ momentum, 
which means that the polarization vector
lies always in the scattering plane,
and (ii) if $m_{\tau}= 0$, then $\vec{s}$ could take only 
$(0,0,\mp\frac{1}{2})$, which means fully left-handed
$\tau^{-}$ or right-handed $\tau^{+}$.

\section{Quasi-Elastic Scattering}\label{secqe}
\hspace*{\parindent}
In this section, we give the spin density matrix calculation
for the QE scattering processes 
\bea
\nu_{\tau} + n \!\!\!&\to&\!\!\! \tau^{-} +p,\\
\bar{\nu}_{\tau} + p \!\!\!&\to&\!\!\! \tau^{+} +n.
\eea
Following Llewellyn Smith\cite{smith}, the hadronic tensor 
is written by using the weak transition current
$J_{\mu}^{(\pm)}$ as follows:
\bea
W^{\rm QE}_{\mu\nu}=\frac{\cos^{2}\theta_{c}}{4}\sum_{\rm spins}
J_{\mu}^{(\pm)}{J^{(\pm)}_{\nu}}^{*}\,\delta(W^2-M^{2}),
\label{qew}\eea
where $\theta_{c}$ is the Cabibbo angle.
The weak transition currents $J_{\mu}^{(+)}$ and $J_{\mu}^{(-)}$ 
for the $\nu_{\tau}$ and $\bar{\nu}_{\tau}$ scattering, 
respectively, are defined as
\bea
&&J_{\mu}^{(+)}=\langle p(p')|\hat{J}_{\mu}^{(+)}|n(p)\rangle = 
\bar{u}_{p}(p')\,\Gamma_{\mu}(p',p)\,u_{n}(p),\\
&&J_{\mu}^{(-)}=\langle n(p')|\hat{J}_{\mu}^{(-)}|p(p)\rangle = 
\bar{u}_{n}(p')\,\overline{\Gamma}_{\mu}(p',p)\,u_{p}(p)=
\langle p(p)|\hat{J}_{\mu}^{(+)}|n(p')\rangle^{*},
\eea
where $\Gamma_{\mu}$ is written in terms of 
the six weak form factors of the nucleon,  
$F^{V}_{1,2,3}$, $F_{A}$, $F_{3}^A$ and $F_p$, as
\bea
\Gamma_{\mu}(p',p)=\gamma_{\mu}\, F^{V}_{1}(q^2) \!\!\!&+&\!\!\!
\frac{i\sigma_{\mu\alpha}q^{\alpha}\xi}{2M}\,F^{V}_{2}(q^2)
+\frac{q_{\mu}}{M}\,F^{V}_{3}(q^2)\no\\
\!\!\!&+&\!\!\!
\left[ \gamma_{\mu}\,F_{A}(q^2)
+\frac{\left(p+p'\right)_{\mu}}{M}\,F^{A}_{3}(q^2)
+\frac{q_{\mu}}{M}\,F_{p}(q^2) \right]\gamma_5.
\eea
For the $\bar{\nu}_{\tau}$ scattering, the vertex 
$\overline{\Gamma}_{\mu}$ is obtained by $\overline{\Gamma}_{\mu}(p',p)
=\gamma_{0}\Gamma_{\mu}^{\dagger}(p,p')\gamma_{0}$. 
We can drop two form factors, $F^V_3$ and $F^A_3$, because of  
time reversal invariance and isospin symmetry (or equivalently 
no second-class currents). Moreover, the vector form factor
 $F_{1}^{V}$ and $F_{2}^{V}$ are related to the 
electromagnetic form factors of nucleons under the conserved 
vector current (CVC) hypothesis:
\bea
F^{V}_{1}(q^{2})= {\displaystyle G^{V}_{E}(q^{2})
-\frac{q^{2}}{4M^{2}}G^{V}_{M}(q^{2}) \over
\displaystyle 1-\frac{q^{2}}{4M^{2}}},\aki
\xi F^{V}_{2}(q^{2})=\frac{\displaystyle G^{V}_{M}(q^{2})-G^{V}_{E}
(q^{2})}{\displaystyle 1-\frac{q^{2}}{4M^{2}}},
\eea
where 
\bea
G^{V}_{E}(q^{2})=\frac{1}{\displaystyle 
\left(1-\frac{q^{2}}{M_{V}^{2}}\right)^{2}},\aki
G^{V}_{M}(q^{2})=\frac{1+\xi}{\displaystyle 
\left(1-\frac{q^{2}}{M_{V}^{2}}\right)^{2}},
\eea
with a vector mass $M_{V}=0.84$ GeV and $\xi=\mu_{p}-\mu_{n}=3.706$. 
$\mu_p$ and $\mu_n$ are the anomalous magnetic moments of proton and 
neutron, respectively. 
For the axial vector form factor $F_{A}$, we adopt the following 
parametrization:
\bea
F_{A}(q^{2})=\frac{F_{A}(0)}{\displaystyle
\left(1-\frac{q^{2}}{M_{A}^{2}}\right)^{2}}
\eea
with an axial-vector mass $M_{A}=1.0$ GeV and $F_{A}(0)=-1.23$\cite{smith}.
For the pseudo-scalar form factor $F_{p}$, 
we adopt the parametrization of Ref.\cite{smith}
\bea
F_{p}(q^{2})=2M^{2}\frac{F_{A}(q^{2})}{m_{\pi}^{2}-q^{2}}\label{fp}
\eea
with the pion mass $m_{\pi}=0.14$ GeV. The normalization 
of $F_{p}(0)$ is fixed by the partially conserved axial vector 
current (PCAC) hypothesis. It should be stressed here that, 
the form factor $F_{p}(q^{2})$ has not been measured experimentally 
because its contribution is proportional to the lepton mass\footnote{
After the paper\cite{taupol} was published, we learned 
that there are some experiments which measure the 
pseudoscalar form factors in muon capture\cite{bardin} 
and in pion electroproduction\cite{choi}. 
Those experiments found results consistent with the 
PCAC relation at very low $Q^2$, but they are not sensitive 
to the large $Q^2$ region ($\approx O(1 {\rm GeV}^2)$) 
which is relevant for $\tau$ production.
A lattice study by Liu {\it et al}.\cite{liu} seems to 
agree with our parametrization of Eq.(\ref{fp}).}.
The production cross section and the polarization of $\tau$ are 
sensitive to $F_{p}(q^{2})$ because of the large $\tau$ mass and the 
 spin-flip nature of the form factor.\\

\begin{figure}[H]
\begin{center}
\epsfig{figure=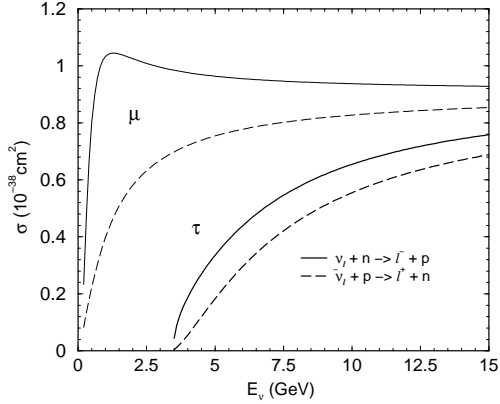,width=6.5cm}
\caption{The neutrino energy dependence of the 
total cross sections of the QE (quasi-elastic) 
processes, $\nu_{l}n \to {l}^{-}p$ (solid lines) and 
$\bar{\nu}_{l}p \to {l}^{+}n$ (dashed lines).
The thick lines are for ${l}=\tau$, while the thin lines are for
${l}=\mu$.}\label{qet}
\end{center}
\end{figure}

In Fig.\ref{qet}, we show the total cross sections of the QE process 
versus the incoming neutrino energy. We plot not only the 
$\tau$-neutrino interaction process, 
but also the $\mu$-neutrino interaction process 
for comparison. Solid curves are the $\nu_{\mu}$ and $\nu_{\tau}$ 
scattering cross sections and dashed curves are the
$\bar{\nu}_{\mu}$ and $\bar{\nu}_{\tau}$ scattering cross sections.  
Our results agree well with those of Hall and Murayama\cite{hm}.

\section{Resonance Production}\label{secres}
\hspace*{\parindent}
In this section, we present the spin density matrix 
calculation for the $\Delta$ production processes
\bea
\nu_{\tau}+n(p)\!\!\!&\to&\!\!\!\tau^{-}+\Delta^{+}(\Delta^{++}),\\
\bar{\nu}_{\tau}+p(n)\!\!\!&\to&\!\!\!
\tau^{+}+\Delta^{0}(\Delta^{-}).\label{REScross}
\eea
We neglect $N^{*}$ and the other higher resonance 
states, which are known to give small contributions\cite{py,sehgal,ppy}. 
For the $\Delta$ resonance production, we calculate 
the hadronic tensor by using the nucleon-$\Delta$ 
weak transition current $J_{\mu}$ as follows:
\bea
W^{\rm RES}_{\mu\nu}=\frac{\cos^{2}\theta_{c}}{4}
\sum_{\rm spins}J_{\mu}J^{*}_{\nu}\,
\frac{1}{\pi}\,\frac{W\Gamma(W)}
{(W^{2}-M_{\Delta}^{2})^{2}+W^{2}\Gamma^{2}(W)}.
\label{reshad}
\eea
Here, $M_{\Delta}$ is the resonance mass, $M_{\Delta}=1.232$GeV,
and $\Gamma(W)$ is its running width estimated by assuming 
the dominance of S-wave $\Delta \to N + \pi$ decay:
\bea
\Gamma(W) =
 \Gamma(M_{\Delta})\,\frac{M_{\Delta}}{W}
\frac{\lambda^{\frac{1}{2}}(W^{2},M^{2},m^{2}_{\pi})}
{\lambda^{\frac{1}{2}}(M_{\Delta}^{2},M^{2},m^{2}_{\pi})}
\eea
with $\Gamma(M_{\Delta})=0.12$ GeV and 
$\lambda(a,b,c)=a^{2}+b^{2}+c^{2}-2(ab+bc+ca)$.\\

The current $J_{\mu}$ for the process 
$\nu_{\tau}+n\to\tau^{-}+\Delta^{+}$ is defined by 
\bea
J_{\mu}=\langle\Delta^{+}(p')|\hat{J}_{\mu}|n(p) \rangle 
=\bar{\psi}^{\alpha}(p')\,\Gamma_{\mu\alpha}\,u_{n}(p),
\eea
where $\psi^{\alpha}$ is the spin-3/2 particle wave function and 
the vertex $\Gamma_{\mu\alpha}$ is expressed in terms of 
the eight weak form factors 
$C^{V,A}_{i=3,4,5,6}$\cite{smith,hippel,fogli,alvarez} as 
\bea
\Gamma_{\mu\alpha}=\left[\frac{C^{V}_{3}}{M}
\left(g_{\mu\alpha}q\dsl-\gamma_{\mu}q_{\alpha}\right)
+\frac{C^{V}_{4}}{M^{2}}\left(g_{\mu\alpha}p'\cdt q
-p'_{\mu}q_{\alpha}\right)\right.\no \\
\left.+\frac{C^{V}_{5}}{M^{2}}\left(g_{\mu\alpha}p\cdt q 
- p_{\mu}q_{\alpha}\right)
+\frac{C^{V}_{6}}{M^{2}}\,q_{\mu}q_{\alpha}
\right]\gamma_{5}\no\\
+\frac{C^{A}_{3}}{M}\left(g_{\mu\alpha}q\dsl
-\gamma_{\mu}q_{\alpha}\right)
+\frac{C^{A}_{4}}{M^{2}}\left(g_{\mu\alpha}p'\cdt q
-p'_{\mu}q_{\alpha}\right)\no\\
+C^{A}_{5}g_{\mu\alpha}+\frac{C^{A}_{6}}{M^{2}}\,q_{\mu}q_{\alpha}.
\eea
By using the isospin invariance and the Wigner-Eckart theorem, 
we obtain the other nucleon-$\Delta$ weak transition currents as 
\bea
\langle\Delta^{++}|\hat{J}_{\mu}|p \rangle = 
\sqrt{3}\langle\Delta^{+}|\hat{J}_{\mu}|n \rangle =
\sqrt{3}\langle\Delta^{0}|\hat{J}_{\mu}|p \rangle =
\langle\Delta^{-}|\hat{J}_{\mu}|n \rangle.\label{WE}
\eea
From the CVC hypothesis, $C^{V}_{6}=0$ and the other vector 
form factors $C^{V}_{i=3,4,5}$ are related to 
the electromagnetic form factors. We adopt the following 
parametrizations:
\bea
C^V_3(q^2)=\frac{C_3^V(0)}
{\displaystyle\left(1-{q^{2}\over M_V^2}\right)^2},\quad 
C^{V}_{4}(q^{2})=-\frac{M}{M_{\Delta}}\,C^{V}_{3}(q^{2}),\quad 
C^{V}_{5}(q^{2})=0,
\eea
with $C_3^V(0)=2.05$ and a vector mass $M_V^2=0.54$ ${\rm GeV^2}$.
For the axial vector form factors $C^{A}_{i=3,4,5}$, 
we use the modified dipole form factors\cite{hippel,alvarez}
\bea
C^{A}_{i=3,4,5}(q^{2}) = C_{i}(0)\left[1-\frac{a_{i}q^{2}}
{b_{i}-q^{2}}\right]\left(1-\frac{q^{2}}{M_{A}^{2}}\right)^{-2}
\eea
with $C_{3}(0)=0$, $C_{4}(0)=-0.3$, $C_{5}(0)=1.2$, 
$a_{4}=a_{5}=-1.21$, $b_{4}=b_{5}=2.0 {\rm GeV}^{2}$ 
and $M_{A}=1.0$ GeV.
And for $C^{A}_{6}$, we use the following relation\cite{singh}:
\bea
C^{A}_{6}(q^{2})=C^{A}_{5}(q^{2})\frac{M^{2}}{m_{\pi}^{2}-q^2},
\label{ca6}\eea
which agrees with the off-diagonal Goldberger-Treiman 
relation in the limit of $m_{\pi}^{2}\to 0$ and $q^{2}\to 0$\cite{hippel}. 
The pseudo-scalar form factor $C_{6}^{A}(q^{2})$ 
has not been measured because its contribution vanishes for massless leptons.
As in the case of the $F_{p}(q^{2})$ form factor of the QE process, 
$C_{6}^{A}(q^{2})$ has significant effects on the $\tau$ production 
cross section and the $\tau$ polarization.\\

In Eq.(\ref{reshad}), summation over the hadronic spins 
is done by using a spin projection operator of the 
spin-3/2 particle wave function which is given by 
\bea
P_{\mu\nu}\!\!\!&=&\!\!\!\sum_{\rm spin}\,\psi_{\mu}(p')\,
\bar{\psi}_{\nu}(p')\no\\
\!\!\!&=&\!\!\!-\left(p\dsl'+M_{\Delta}\right)\left\{g_{\mu\nu}
-\frac{2}{3}\frac{p'_{\mu}p'_{\nu}}{M_{\Delta}^{2}}+\frac{1}{3}
\frac{p_{\mu}'\gamma_{\nu}-p_{\nu}'\gamma_{\mu}}{M_{\Delta}}
-\frac{1}{3}\gamma_{\mu}\gamma_{\nu}\right\}.
\eea
The hadronic tensor is now calculated by
\bea
W^{\rm RES}_{\mu\nu}=\frac{\cos^{2}\theta_{c}}{4}\,
{\rm Tr}\left[P^{\beta\alpha}\Gamma_{\mu\alpha}(p\dsl+M)
\overline{\Gamma}_{\nu\beta}\right]
\frac{1}{\pi}\,\frac{W\Gamma(W)}
{(W^{2}-M_{\Delta}^{2})^{2}+W^{2}\Gamma^{2}(W)}.
\eea\\

\begin{figure}[H]
\begin{center}
\epsfig{figure=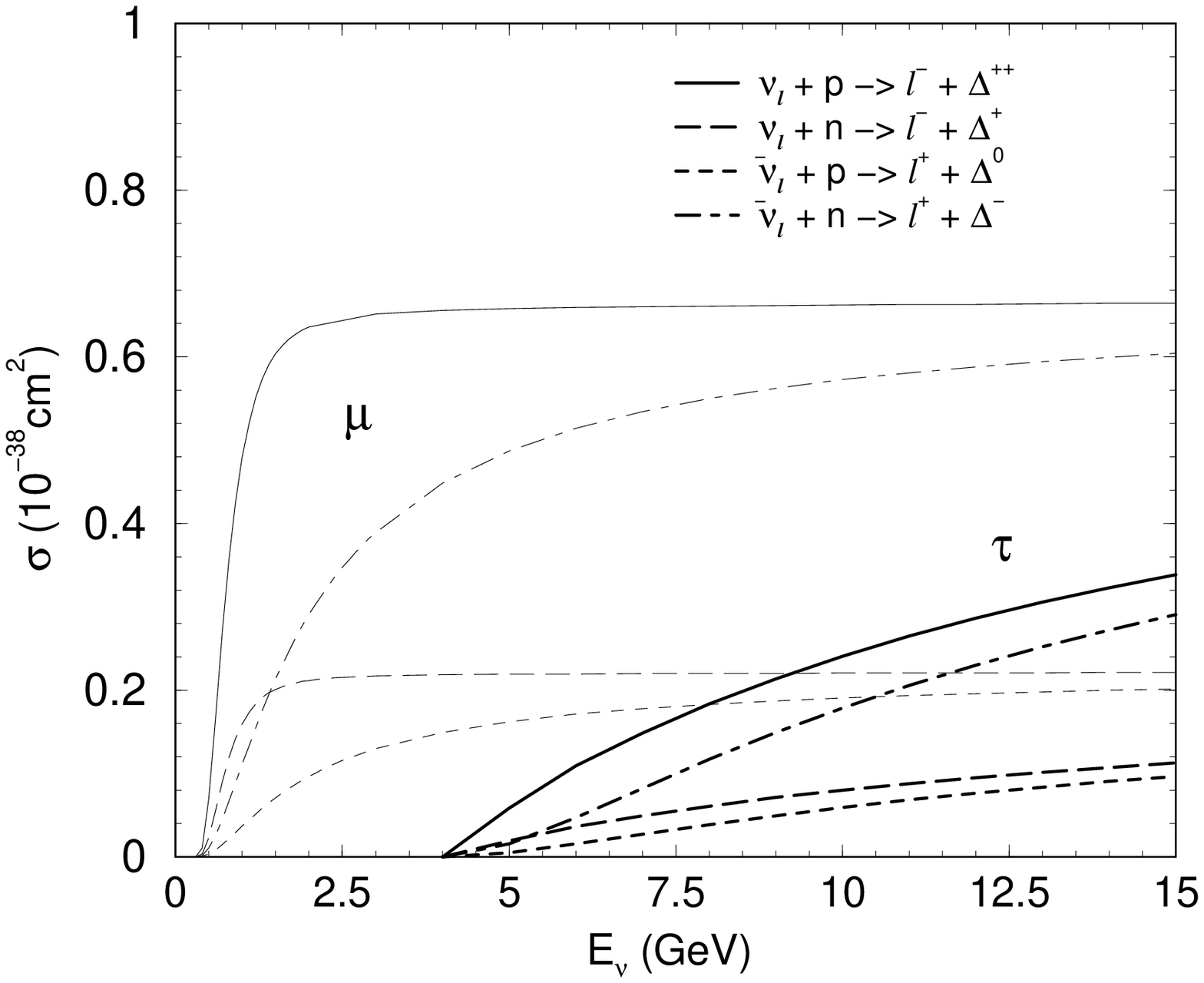,width=6.5cm}
\caption{Total cross sections of the $\Delta$ production (RES) 
processes, plotted against the incoming (anti)neutrino energy. 
The solid, long-dashed, dashed, and dot-dashed lines 
show $\Delta^{++}$, $\Delta^{+}$, $\Delta^{0}$, 
and $\Delta^{-}$ production cross sections, 
respectively. 
The thick lines are for $\nu_{\tau}$ and $\bar{\nu}_{\tau}$ scatterings
 and the thin lines are for $\nu_{\mu}$ and $\bar{\nu}_{\mu}$ scatterings.}
\label{rest}
\end{center}
\end{figure}

By integrating over $E_\tau$ and $\cos\theta$ within the 
kinematical region of $M + m_\pi < W < 1.4$GeV,
we estimate the total cross section of the $\Delta$ 
production (RES) processes. 
In Fig.\ref{rest}, we show the total cross section 
versus the incoming neutrino energy. 
We also plot the total $\Delta$ production cross sections
 for $\nu_{\mu}$ and $\bar{\nu}_{\mu}$ scattering processes, 
in order to examine the lepton mass dependence. The $\mu^{\pm}$ 
production cross sections grow sharply at low $E_{\nu}$,  
 while the $\tau$ production cross section grow mildly  
from around $E_{\nu}=4$GeV. The cross sections of 
$\Delta^{++}$ and $\Delta^{-}$ production processes are larger than 
those of $\Delta^{+}$ and $\Delta^{0}$ productions. This feature 
is expected from the Clebsh-Gordan coefficients of 
the transition currents, in Eq.(\ref{WE}). Our results agree 
approximately with those of Paschos and Yu\cite{py}, which include 
the contributions from $N^{*}(S_{11},P_{11})$ resonance productions.

\section{Deep Inelastic Scattering}\label{secdis}
\hspace*{\parindent}
In this section, we present the spin density matrix calculation 
for the deep inelastic scattering (DIS) processes
\bea
\nu_{\tau} + N \!\!\!&\to&\!\!\! \tau^{-} + X,\\
\bar{\nu}_{\tau} + N \!\!\!&\to&\!\!\! \tau^{+} + X.
\eea
In the DIS region, the hadronic tensor is estimated by using the
quark-parton model;
\bea
W^{\rm DIS}_{\mu\nu}(p,q)=\sum_{q,\bar{q}}\int\frac{d\xi}{\xi}\,
f_{q,\bar{q}}(\xi,Q^{2})\,
K^{(q,\bar{q})}_{\mu\nu}(p_{q},q).\label{disten}
\eea
Here, $p_{q}^{\mu}=\xi p^{\mu}$ is the four-momentum 
of the scattering quark, 
$\xi$ is its momentum fraction, and  
$f_{q}$ and $f_{\bar{q}}$ are the parton distribution function(PDF)'s 
inside a nucleon. By taking the spin average of initial quark and 
by summing over the final quark spins, we find the quark tensor 
\bea
K^{(q,\bar{q})}_{\mu\nu}(p_{q},q)\!\!\!&=&\!\!\!
\delta(2\,p_{q}\cdt q-Q^{2}-m_{q'}^{2})\no \\
&&\times 2\left[-g_{\mu\nu}(p_{q}\cdt q)+2p_{q\mu}p_{q\nu}
\mp i\epsilon_{\mu\nu\alpha\beta}p_{q}^{\alpha}q^{\beta}
+(p_{q\mu}q_{\nu}+q_{\mu}p_{q\nu})\right]. \label{quaten}
\eea
The upper sign should be taken for quarks and the lower for antiquarks. 
We retain the final quark mass, $m_{q'}$, for the charm quark as  
$m_{c}=1.5$GeV, but otherwise we set $m_{q'}=0$. 
We neglect charm and heavier-quark distributions in the nucleon,
as well as bottom and top production cross sections.\\

By neglecting the nucleon mass and the initial quark masses consistently,  
we find the following relations: 
\bea
W_{1}(p\cdt q,Q^{2})=F_{1}(x,Q^{2}),
\aki  W_{i=2,\ldots,5}(p\cdt q,Q^{2})= 
{M^2 \over p\cdot q }\,  F_{i=2,\ldots,5}(x,Q^{2}).
\eea
Here,
\begin{subequations}
\bea
&&F_{1}=\sum_{q,\bar{q}}f_{q,\bar{q}}(\xi,Q^{2}),\\
&&F_{2}=2\sum_{q,\bar{q}}\xi\,f_{q,\bar{q}}(\xi,Q^{2}),\\
&&F_{3}=2\sum_{q}f_{q}(\xi,Q^{2})
-2\sum_{\bar{q}}f_{\bar{q}}(\xi,Q^{2}),\\
&&F_{4}=0\,,\\ 
&&F_{5}=2\sum_{q,\bar{q}}f_{q,\bar{q}}(\xi,Q^{2}),  
\eea
\end{subequations}
where the momentum fraction is $\xi = x$ for massless final quarks 
($m_{q'}=0$), and  
$\xi = {x}/{\lambda}$ with $\lambda = Q^{2}/(Q^{2}+m_{q'}^{2})$ for $q'=c$.
In the $m_c\to 0$ limit, the 
Callan-Gross relation $F_{2}=2xF_{1}$ and the  
Albright-Jarlskog relations $F_{4}=0$, $2xF_{5}=F_{2}$ hold.\\

However, the differential cross section (Eq.(\ref{cross})) does not satisfy 
the positivity condition near the threshold with this naive replacement.
We find that the following modification of the $W_1$ structure function 
suffices to ensure the positivity constraints\footnote{Slightly more 
complicated rescaling low has been examined by 
Albright and Jarlskog\cite{albright}.}:
\be
W_1= \left(1+{\xi M^2 \over p\cdot q}\right)F_1, 
\label{W1}
\ee
and find that the positivity is maintained when the charm quark mass is 
introduced by using the rescaling variable $\xi = {x}/{\lambda}$.\\

There is further uncertainty in our parton model predictions for the inelastic 
scattering processes where the hadronic final state is heavy, $W\geq 1.4$GeV, 
but the momentum transfer is small, $Q^2\leq 1$GeV$^2$. This is the region of 
the phase space depicted by the star symbol ($\ast$) in the $x$-$y$ plane  
and the $p_{\tau}\cos\theta$-$p_{\tau}\sin\theta$ plane of 
Fig.\ref{kinema10}. In order to estimate the cross section and the spin 
polarization vector in this region, we use naive extrapolation of the parton 
model calculation, by using the parton distribution at the minimum 
$Q^2$ ($Q^2_0=1.25$GeV$^2$ for the parametrization of 
A.D.Martin {\it et al}.\cite{mrst}) even when $Q^2 < Q^2_0$. \\

\begin{figure}[H]
\begin{center}
\epsfig{figure=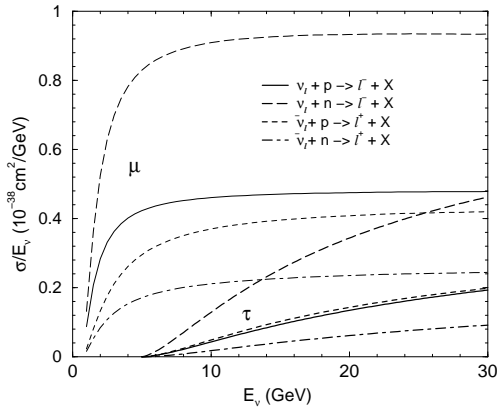,width=6.5cm}
\caption{Total cross sections of the DIS processes divided by the  
neutrino energy are plotted against the neutrino energy. 
Solid, long-dashed, dashed, and dot-dashed 
lines show $\nu_{l}p \to l^{-}X$, $\nu_{l}n \to l^{-}X$, 
$\bar{\nu}_{l}p \to l^{+}X$, and $\bar{\nu}_{l}n \to l^{+}X$ 
processes, respectively. 
The thick lines are for $l=\tau$ and the thin lines are for $l=\mu$.
}\label{dist}
\end{center}
\end{figure}

In Fig.\ref{dist}, we plot the total cross sections of the  
DIS (IS) process for $\nu_{\tau}N$ and $\bar{\nu}_{\tau}N$ scatterings by 
thick lines. Those of the $\nu_{\mu}N$ and $\bar{\nu}_{\mu}N$ scattering 
processes are shown by thin lines for comparison. Those curves are obtained by 
using the parton distribution function(PDF)'s of Martin 
{\it et al}.\cite{mrst}. 
The results are similar to the RES case, $\mu$ production 
cross sections grow rapidly from low $E_{\nu}$, 
and the $\tau$ production cross sections grow mildly from around  
$E_{\nu}=5$GeV. These results are consistent with the calculations
of Kretzer and Reno\cite{kretzer}, which include the NLO corrections.\\

Uncertainties in the total cross section due to the modification of the 
structure function $W_1$ (Eq.(\ref{W1})) and in the contribution from the 
$Q^2 < 1$GeV$^2$ region are found to be rather small. A more quantitative 
study of the uncertainty in the theoretical predictions will be reported 
elsewhere. \\

\begin{figure}[H]
\begin{center}
\epsfig{figure=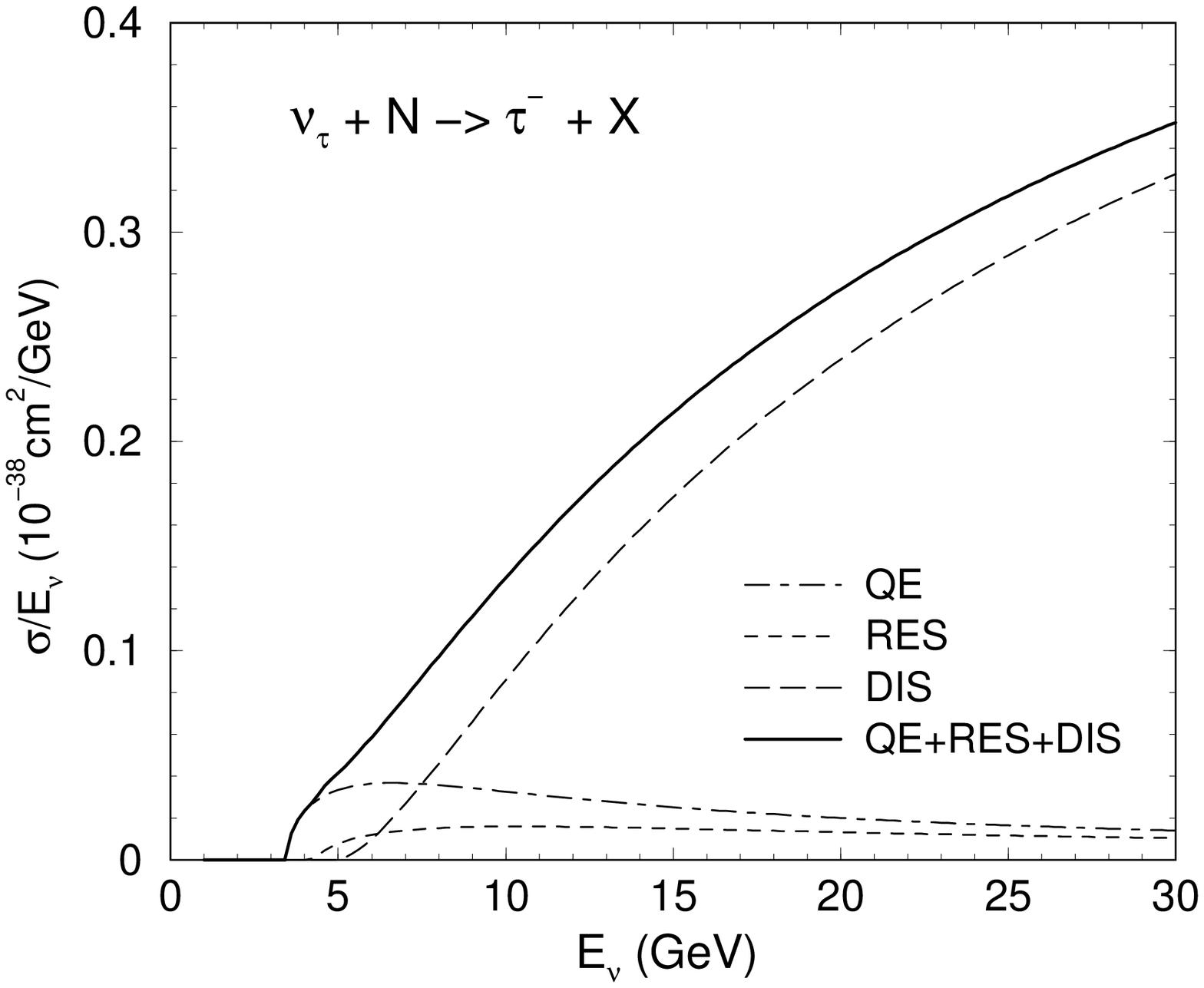,width=6.5cm}
\aki\epsfig{figure=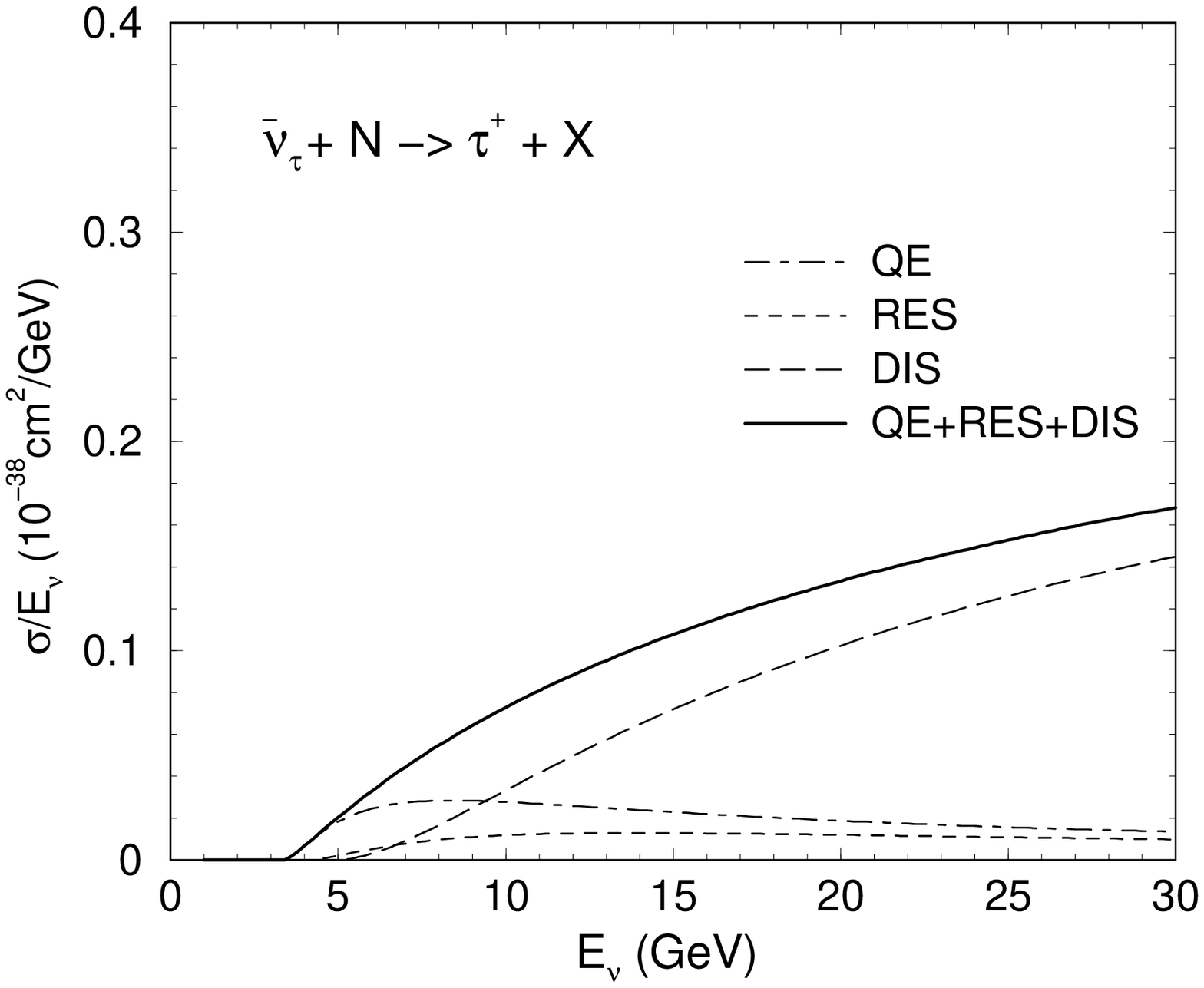,width=6.5cm}
\caption{The neutrino energy dependence of the total cross section of 
$\tau^{-}$ (left) and $\tau^{+}$ (right) productions 
off the isoscalar target, normalized by incoming neutrino energy. 
The contributions from QE, RES and DIS processes are shown by  
dot-dashed, dashed and long-dashed lines, respectively, and their sums 
are shown by thick solid lines. 
}\label{total}
\end{center}
\end{figure}

In Fig.\ref{total}, we show the total cross section 
of all the $\tau^{\pm}$ production process 
for the isoscalar target.
The cross sections normalized to the neutrino energy are plotted 
against the neutrino energy.
The left figure is for $\tau^-$ production and the right figure is 
for $\tau^+$ production. We find that at medium neutrino energies, 
the QE contribution dominates the total cross section near the threshold,
and the sum of the QE and RES cross sections are significant throughout 
the energy range of the future neutrino oscillation experiments. 
Significance of the QE and RES contribution is more pronounced for the 
$\bar{\nu}_{\tau}N \to \tau^{+}X$ reaction shown in the right-hand figure, 
where the DIS contribution starts dominating the total cross section 
only above $E_{{\nu}}=10$GeV. Those trends agree with the earlier 
results of Paschos and Yu\cite{py}.

\section{Polarization of the produced $\tau^{\pm}$}\label{result}
\hspace*{\parindent}
In this section, we show the spin polarization vector 
of the produced $\tau$ lepton as a function of its 
energy $E_{\tau}$ and the scattering angle $\theta$ in the laboratory frame. 
We show our results for two arbitrarily fixed neutrino and antineutrino 
energies, $E_{\nu}=10$GeV and $20$GeV, for isoscalar targets.\\

\begin{figure}[H]
\begin{minipage}{5cm}
\epsfig{figure=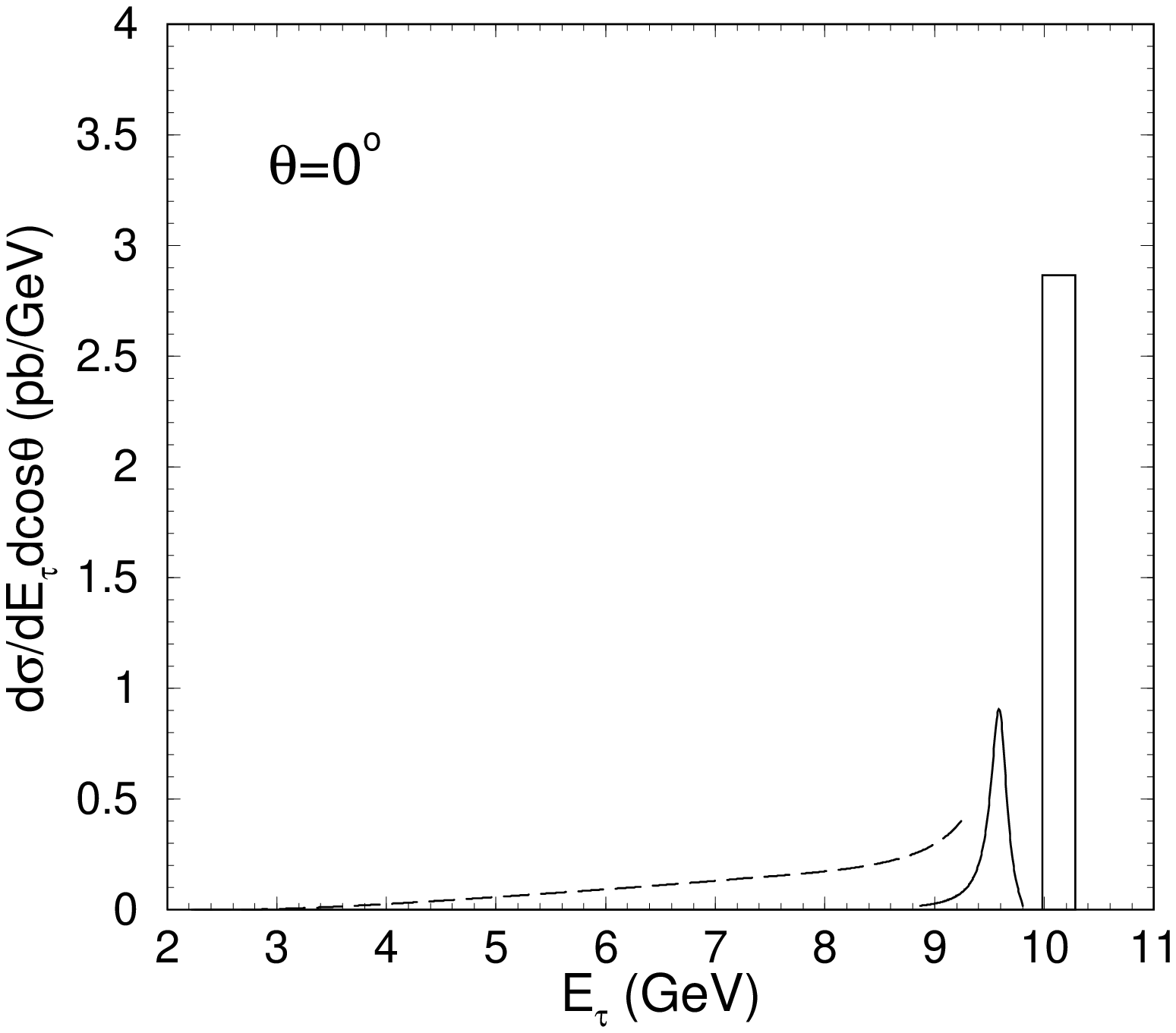,width=5cm}
\end{minipage}
\begin{minipage}{5cm}
\epsfig{figure=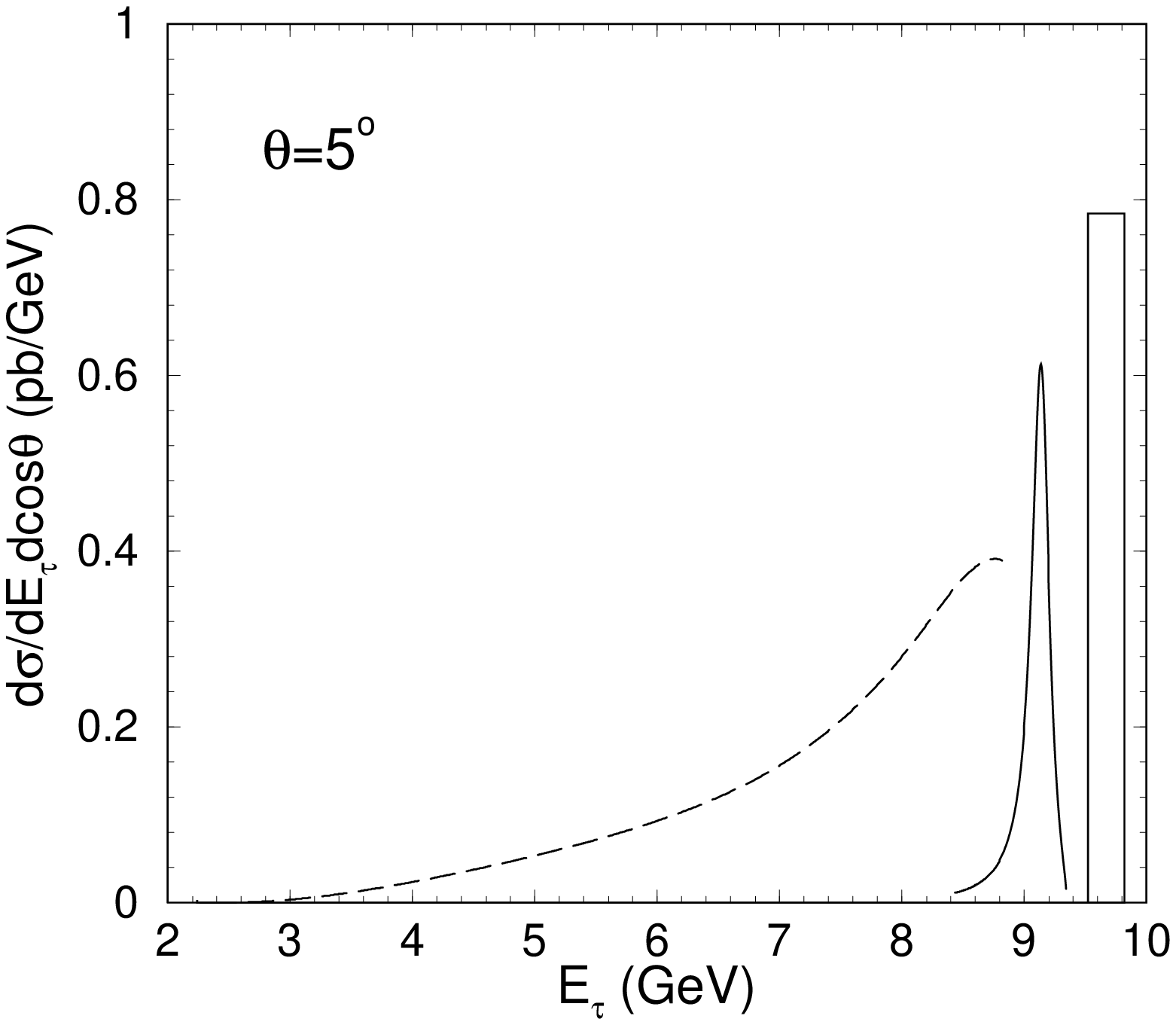,width=5cm}
\end{minipage}
\begin{minipage}{5cm}
\epsfig{figure=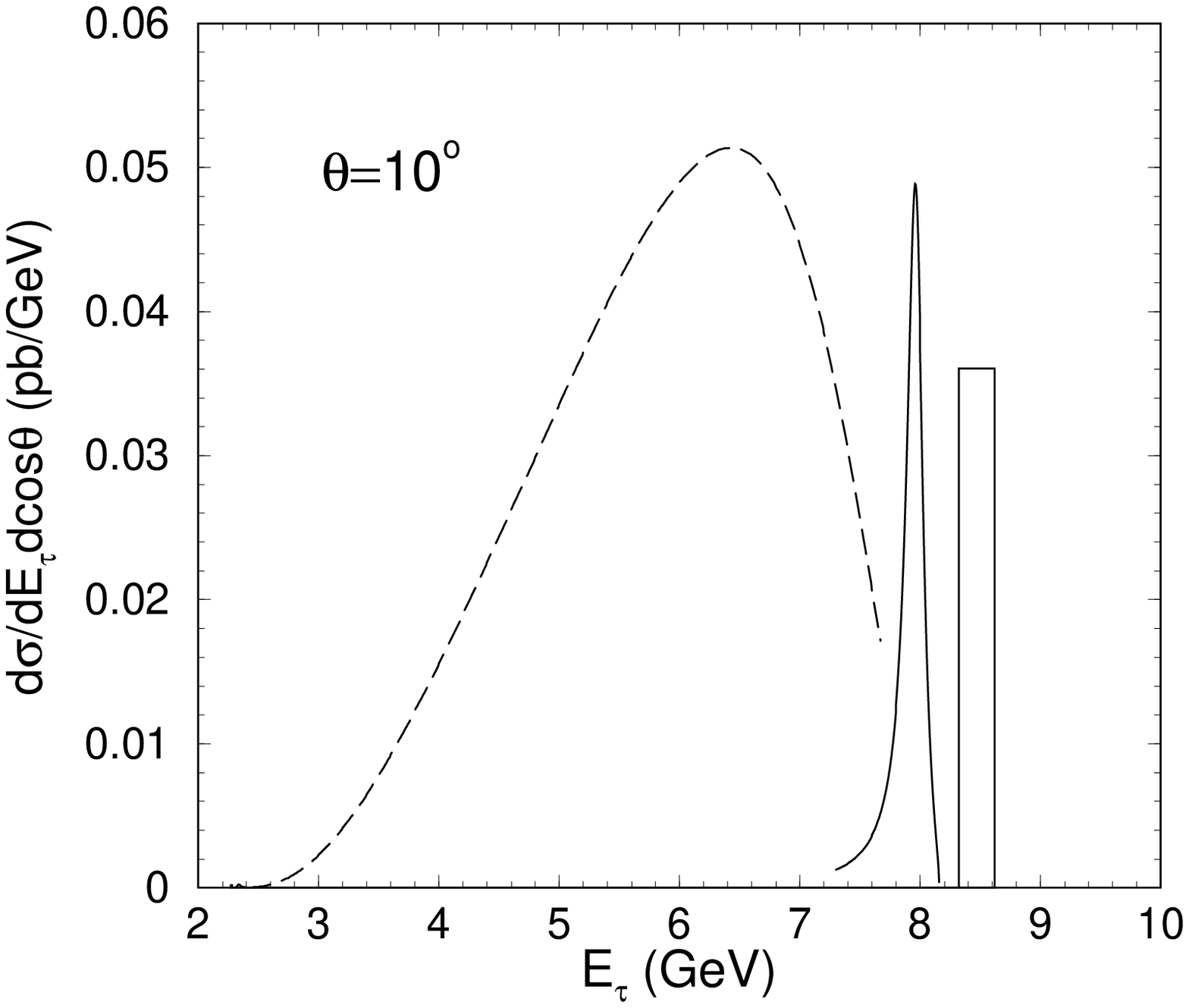,width=5cm}
\end{minipage}\\
\begin{minipage}{5cm}
\epsfig{figure=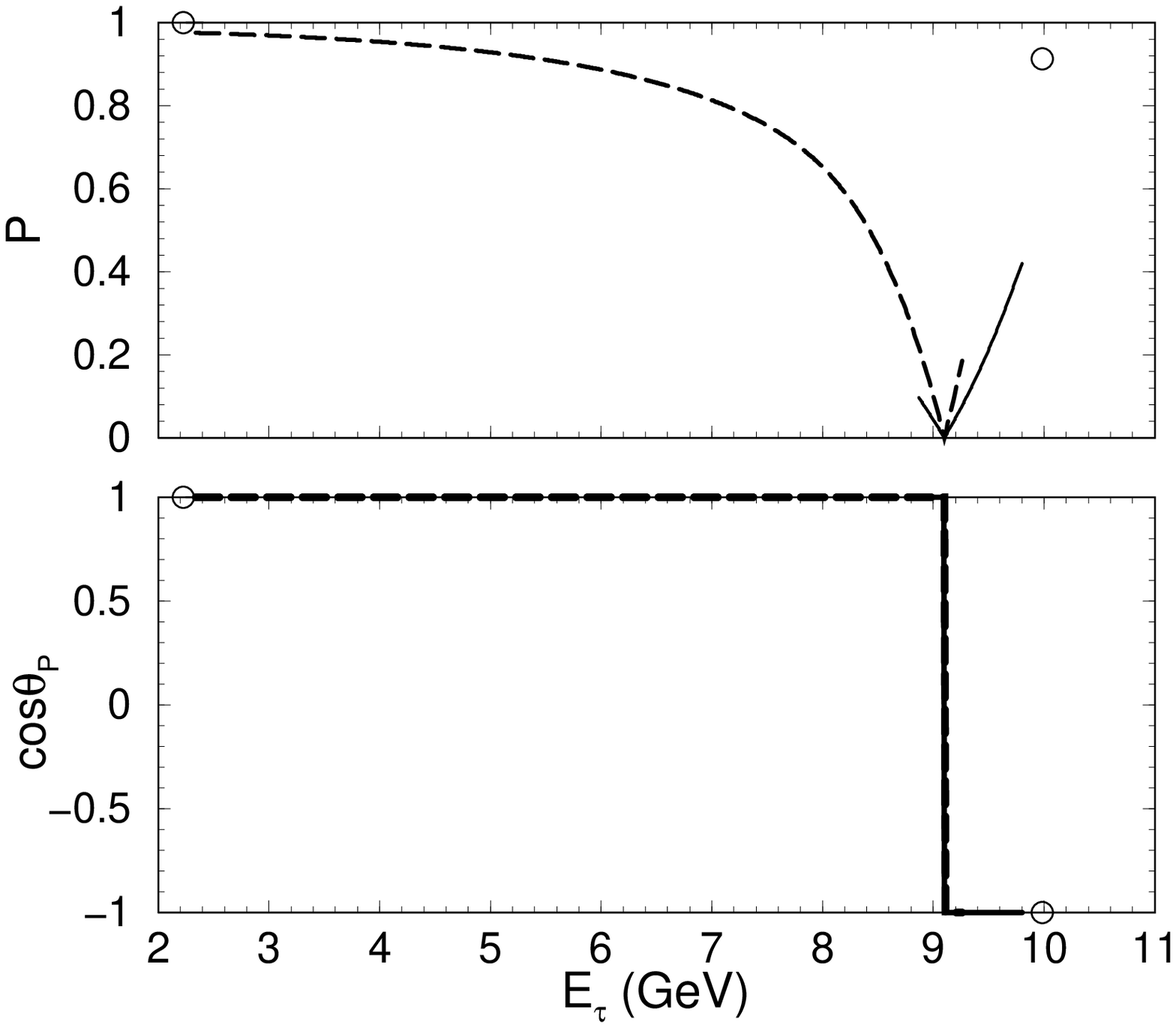,width=5cm}
\end{minipage}
\begin{minipage}{5cm}
\epsfig{figure=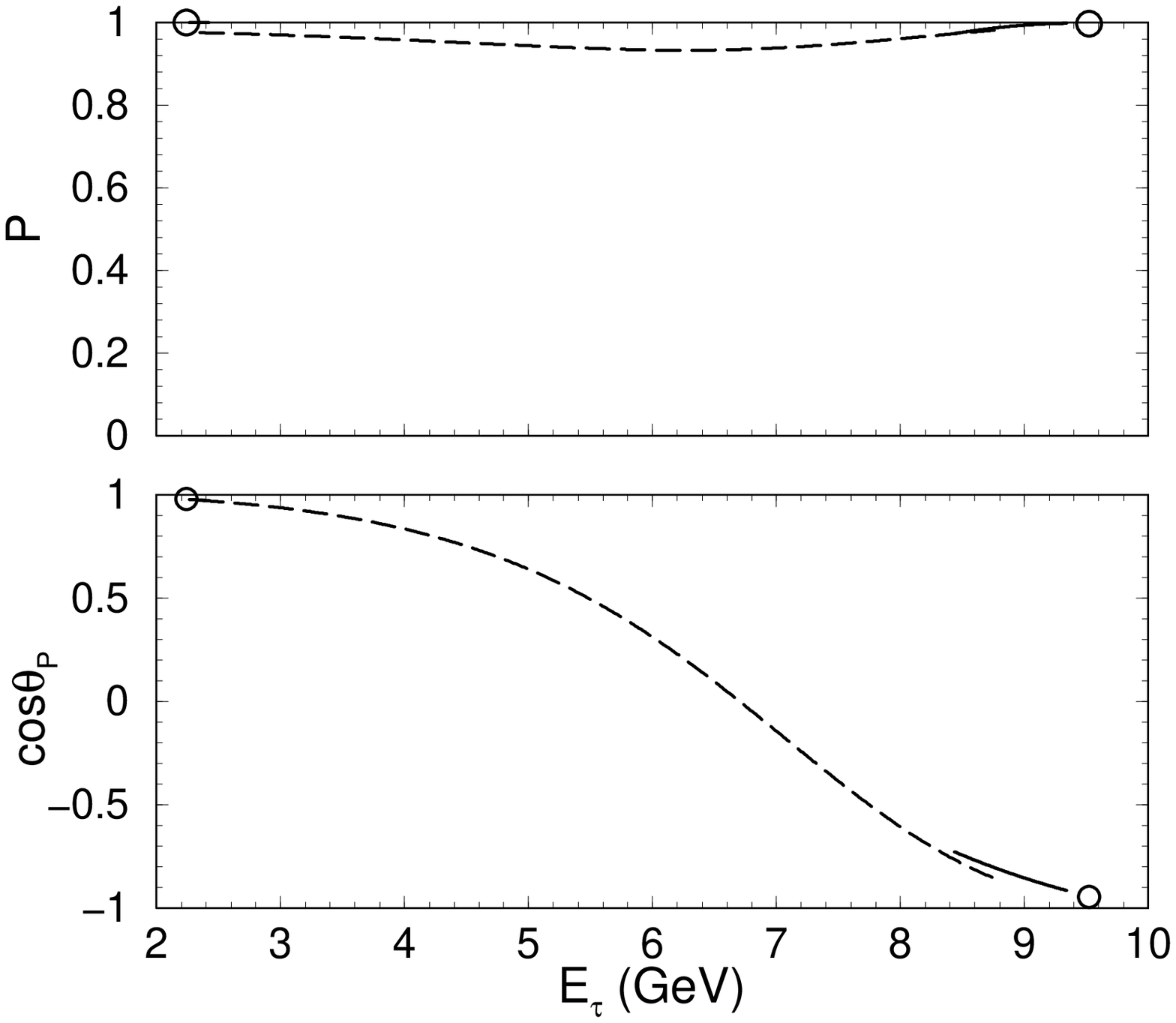,width=5cm}
\end{minipage}
\begin{minipage}{5cm}
\epsfig{figure=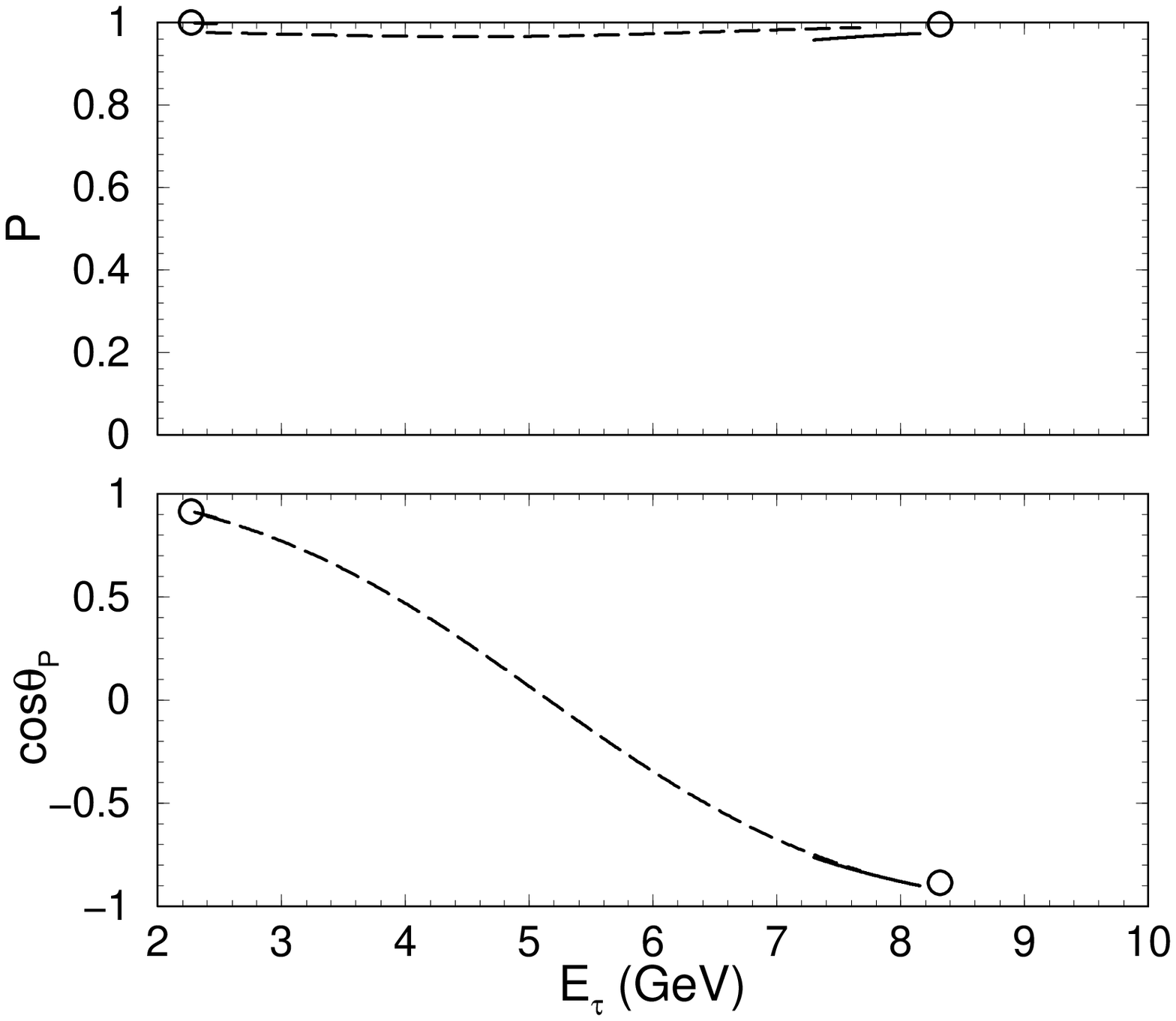,width=5cm}
\end{minipage}
\caption{
Production cross section and the $\tau$ polarization of the process 
${\nu}_{\tau}N \to \tau^{-}X$ at $E_{\nu}=10$GeV. $E_{\tau}$ dependence 
of the differential cross section (top), the degree of 
polarization $P$ (middle) and the polar component of the normalized 
polarization vector $\cos\theta_P$ (bottom) are shown along the laboratory 
frame scattering angle $\theta=0^{\circ}$ (left), $5^{\circ}$ (center) and
$10^{\circ}$ (right), respectively. The  
Histograms in the top figures and the circles in the middle and the bottom 
figures represent QE process, solid lines show RES process, and the 
dashed lines are for DIS process. The $\tau^-$ spin in the $\tau^-$ rest 
frame is $\vec{s}={P\over 2}(-\sin\theta_P,0,\cos\theta_P)$.
}\label{cross10}
\end{figure}

Fig.\ref{cross10} summarizes our results for the 
${\nu}_{\tau}N \to \tau^{-}X$ process at $E_{\nu}=10$GeV. 
The top three figures show the double differential cross 
section, Eq.(\ref{cross}), as a function of $E_{\tau}$, at 
$\theta=0^{\circ}$ (left figures), $5^{\circ}$ (center figures) and
$10^{\circ}$ (right figures). The DIS (IS) contributions are shown by dashed 
lines, and the RES and QE contributions are shown by the solid lines. 
The area of the histogram for the QE process is normalized to the cross 
section. A set of three middle figures give the degree of polarization, 
$P$ of Eq.(\ref{spinvector}), as functions of $E_{\tau}$.
In the bottom three figures, we show the $E_{\tau}$ dependence of the 
polarization direction, Eq.(\ref{spinvector}), by using 
$\cos\theta_P=s_z/{(P/2)}$. This suffices to determine the polarization 
direction because $s_x$ turns out to be always negative, 
$s_x=-{P\over 2}\sin\theta_P$, and $s_y=0$ ($\varphi_P=\pi$). 
It should be noted that all 
the 9 figures have common horizontal scale. The overall phase space of 
the $E_{\nu}=10$GeV experiment in the laboratory frame has been shown 
in Fig.\ref{kinema10}(right). \\

The differential cross sections are obtained from Eq.(\ref{cross}). 
According to the phase-space plot of Fig.\ref{kinema10}(right), 
along a fixed laboratory scattering angle $\theta$, there are two 
$E_{\tau}$'s at which QE and RES reactions can take place. 
The top figures of Fig.\ref{cross10} show us that the cross sections 
in the lower $E_{\tau}$ sides are negligible.
This is because of the form factor suppression which is significant 
already at $E_{\nu}=10$GeV. The QE and RES cross sections are large at 
forward scattering angles, and the DIS contribution become more significant 
at large scattering angles, though the cross section gets smaller. 
In order to examine the transition between the $\Delta$ resonance 
production (RES) process and the DIS process, we show our predictions for RES up to 
$W<1.6$GeV and those for DIS from $W>1.4$GeV, allowing for the overlap. 
Although there is no strong reason to expect smooth transition, we 
find the tendency that our predictions for the $\tau^{\pm}$ production 
are relatively smoothly changing in the transition region.\\

The degree of polarization $P$ and the polar angle 
$\theta_P$ are defined in Eq.(\ref{spinvector}). 
The produced $\tau^{-}$ is almost fully polarized except at 
 the very small scattering angle. 
As for the angle of the polarization vector, 
the high energy $\tau^{-}$ is almost left-handed ($\cos\theta_P=-1$). 
On the other hand, the spin of low energy $\tau^{-}$ turns around. 
The azimuthal angle $\varphi_{P}$ takes $\pi$ at all energies, 
which means that the spin vector points to the direction of 
the initial neutrino momentum axis.\\

In order to understand the above features, it is useful to consider the
polarization of $\tau^-$ in the center of mass (CM) frame of the 
scattering particles. Let us consider the DIS process in the $\nu q$ 
CM frame, since the $\nu q$ scattering is dominant in the 
$\nu_{\tau}N\to\tau^{-}X$ process. In this frame, 
produced $\tau^{-}$ is fully left-handed polarized at all scattering angles. 
This is because the initial $\nu_{\tau}$ and $q$ ($d$ or $s$ quarks) are 
both left-handed, and hence angular momentum along the initial momentum 
direction is zero, while in the final state, the produced $u$ quark is 
left-handed and hence only the left-handed $\tau^-$ is allowed by the angular 
momentum conservation. This selection rule is violated slightly when a 
charm quark is produced in the final state and because of gluon radiation 
at higher orders of QCD perturbation theory. The $\tau^-$ polarization in the 
laboratory frame is then obtained by the Lorentz boost.
In the QE and RES processes, situations are
almost the same as in the DIS process. In the CM frame of $\nu N$ collisions, 
the $\tau^-$ lepton produced by the QE or RES process is almost 
left-handed at all angles, for the CM energy of 
$\sqrt{2ME_{\nu}+M^2}\approx 4.4$GeV for $E_{\nu}=10$GeV, for our 
parametrizations of the transition form factors.
High energy $\tau^-$'s in the laboratory frame have 
left-handed polarization because those $\tau^-$'s have forward scattering 
angles also in the CM frame.
However, lower energy $\tau^-$'s in the laboratory frame tends to have 
right-handed polarization because they are produced at 
backward angles in the CM frame. 
At the zero scattering angle $\theta=0^{\circ}$ of the laboratory frame, 
the change in the $\tau^-$ momentum direction occurs suddenly, 
and hence the transition from the left-handed $\tau^-$ at high energies 
to the right-handed $\tau^-$ at low energies is discontinuous.
Since the degree of polarization $P$ vanishes at this point, the 
polarization vector, or the density matrices are continuous. \\

\begin{figure}[H]
\begin{minipage}{5cm}
\epsfig{figure=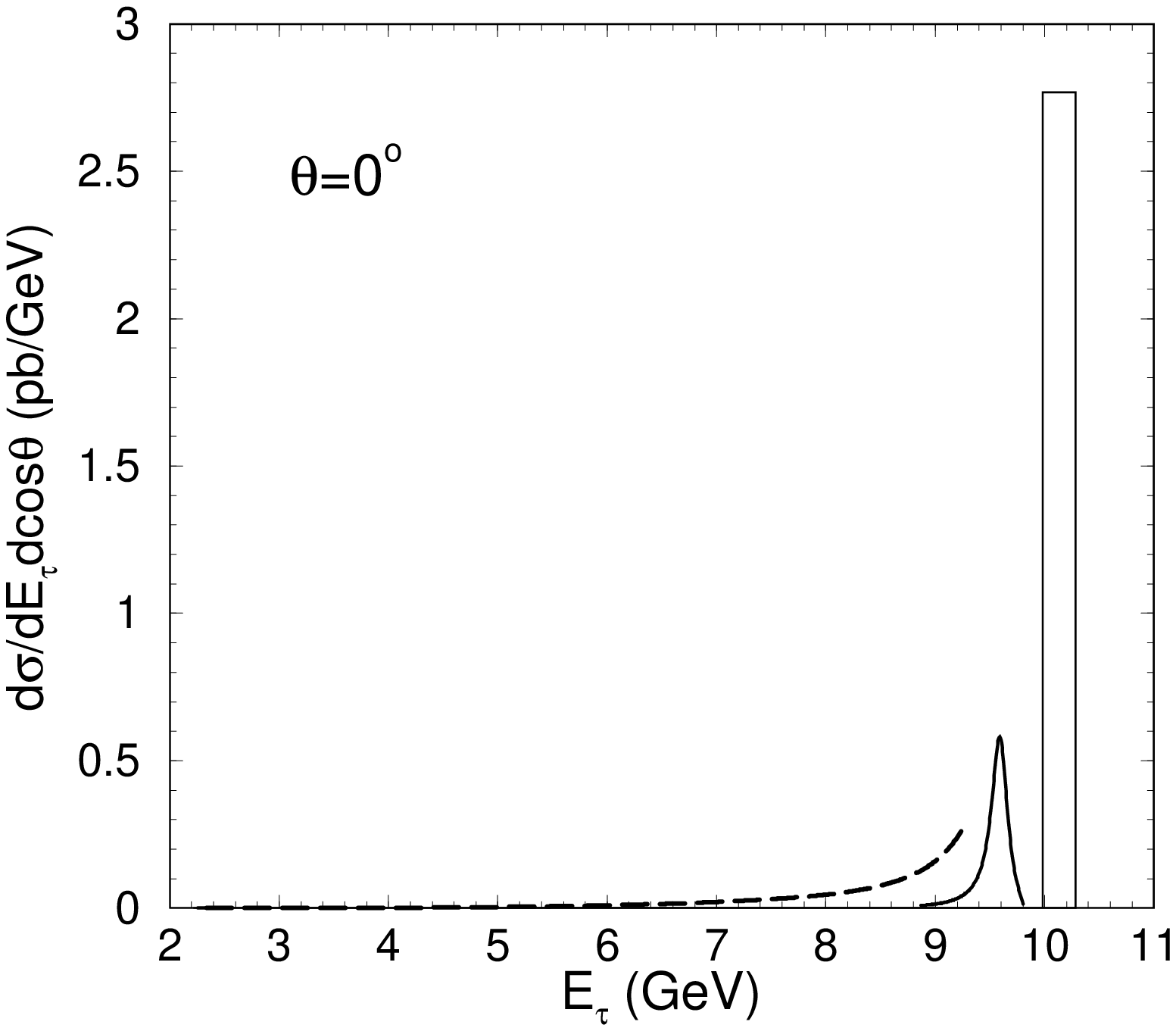,width=5cm}
\end{minipage}
\begin{minipage}{5cm}
\epsfig{figure=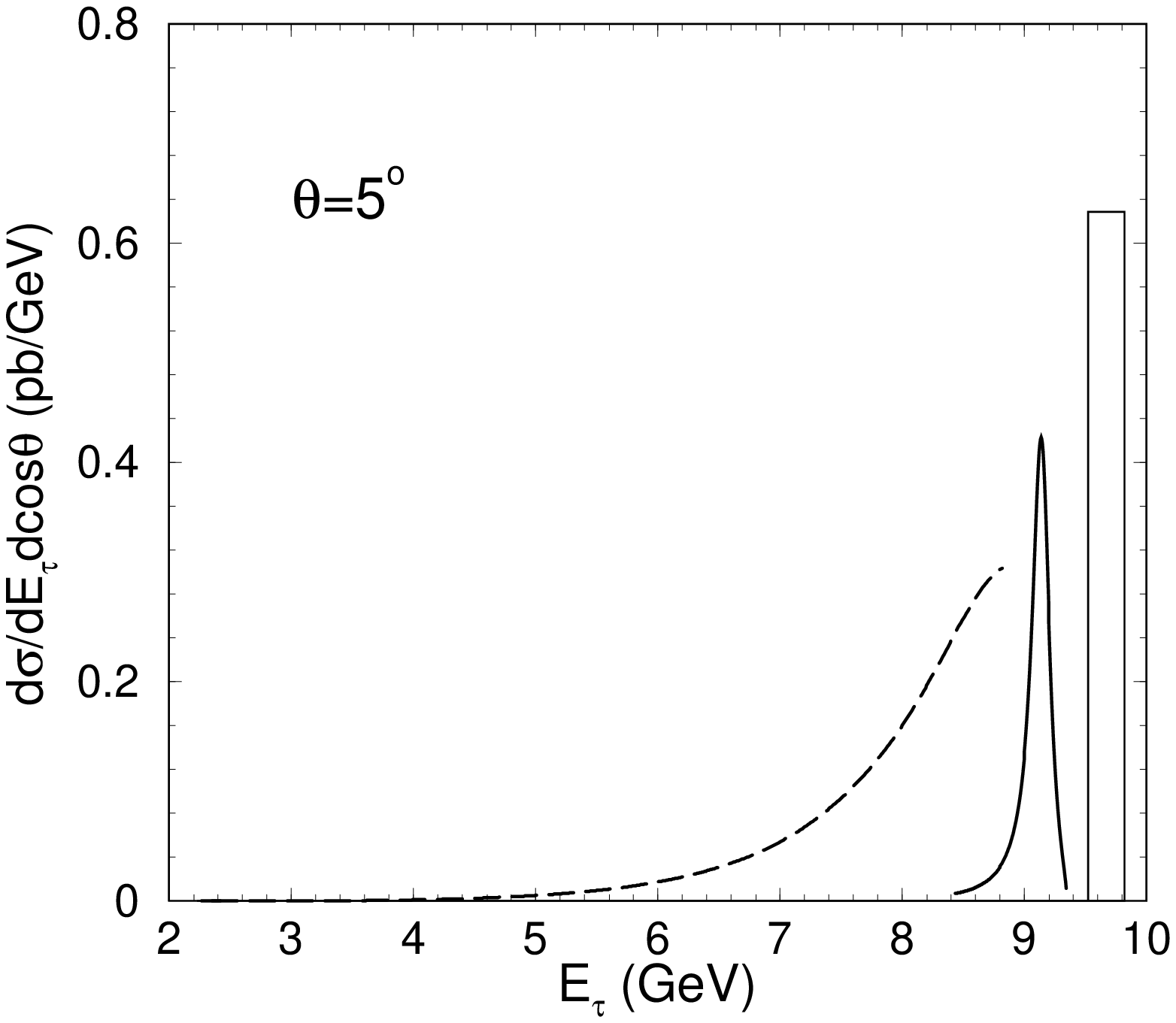,width=5cm}
\end{minipage}
\begin{minipage}{5cm}
\epsfig{figure=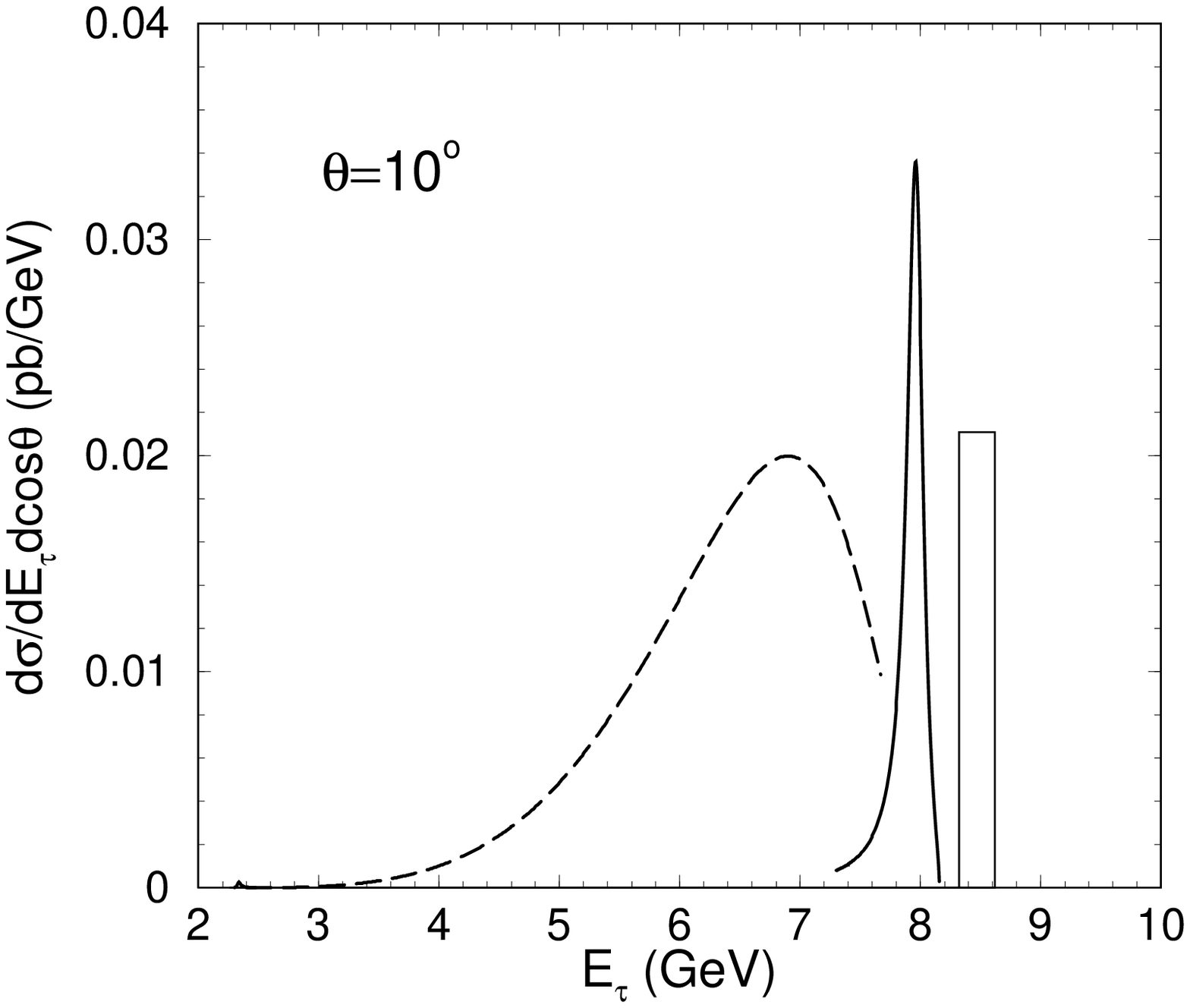,width=5cm}
\end{minipage}\\
\begin{minipage}{5cm}
\epsfig{figure=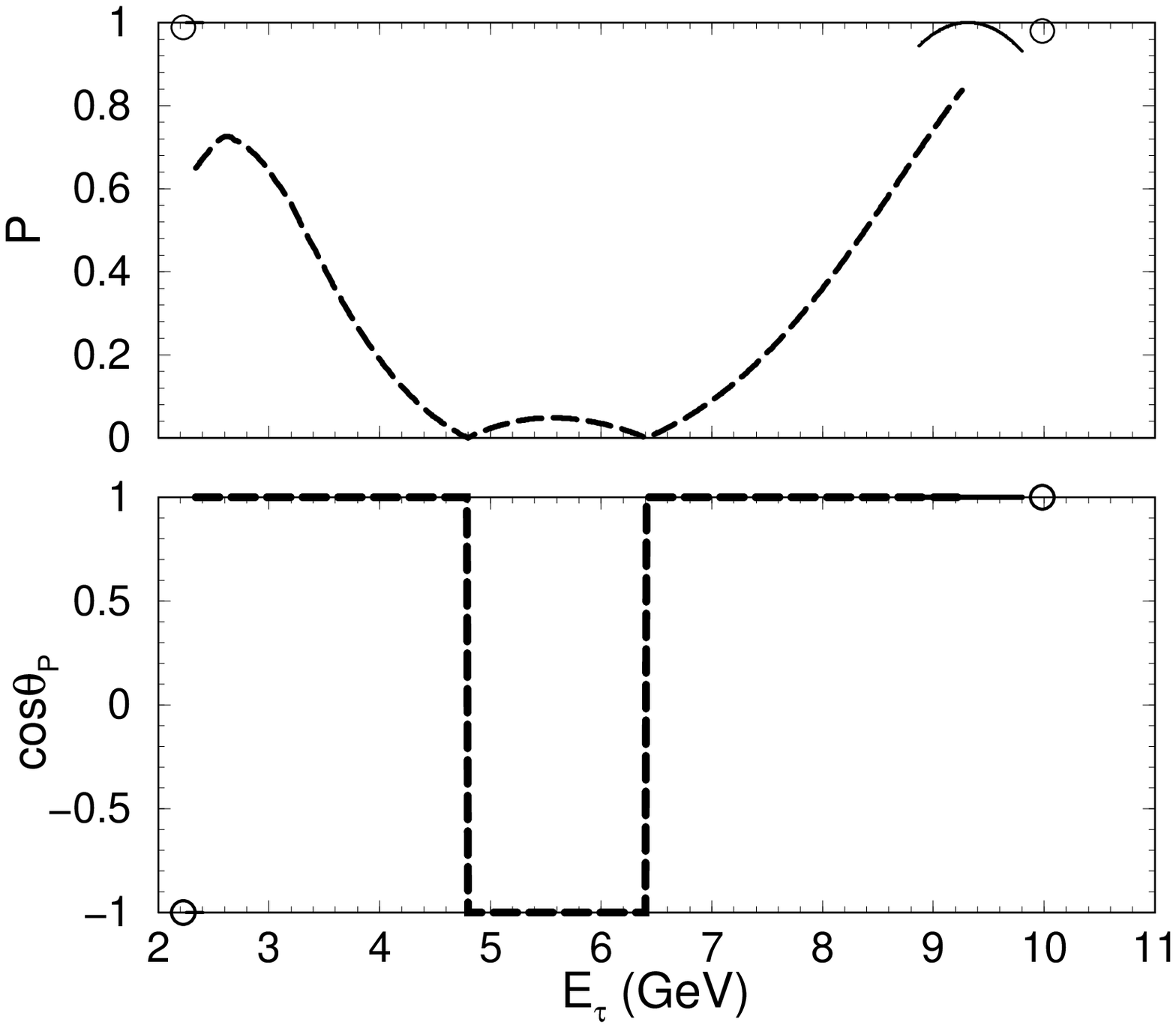,width=5cm}
\end{minipage}
\begin{minipage}{5cm}
\epsfig{figure=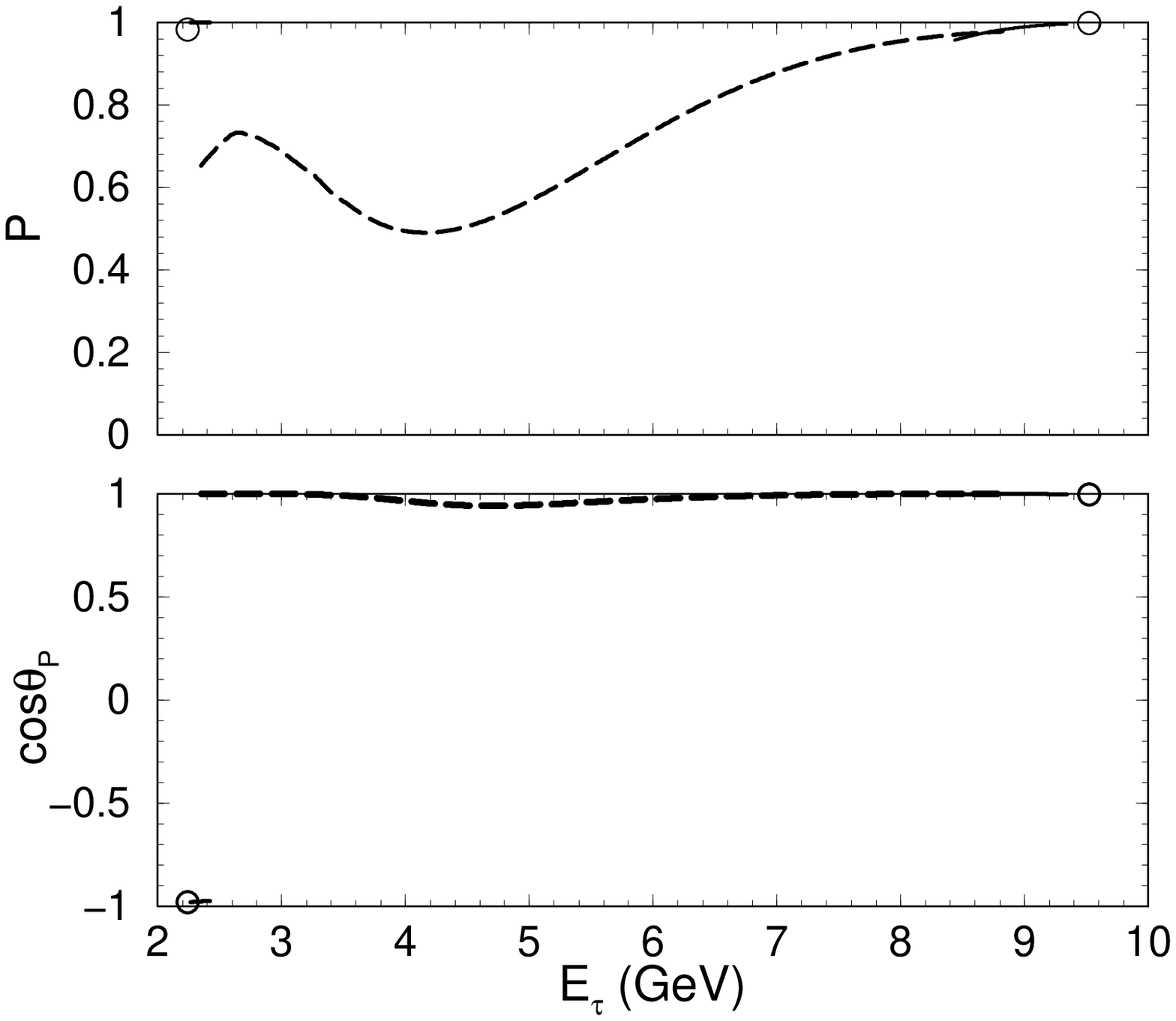,width=5cm}
\end{minipage}
\begin{minipage}{5cm}
\epsfig{figure=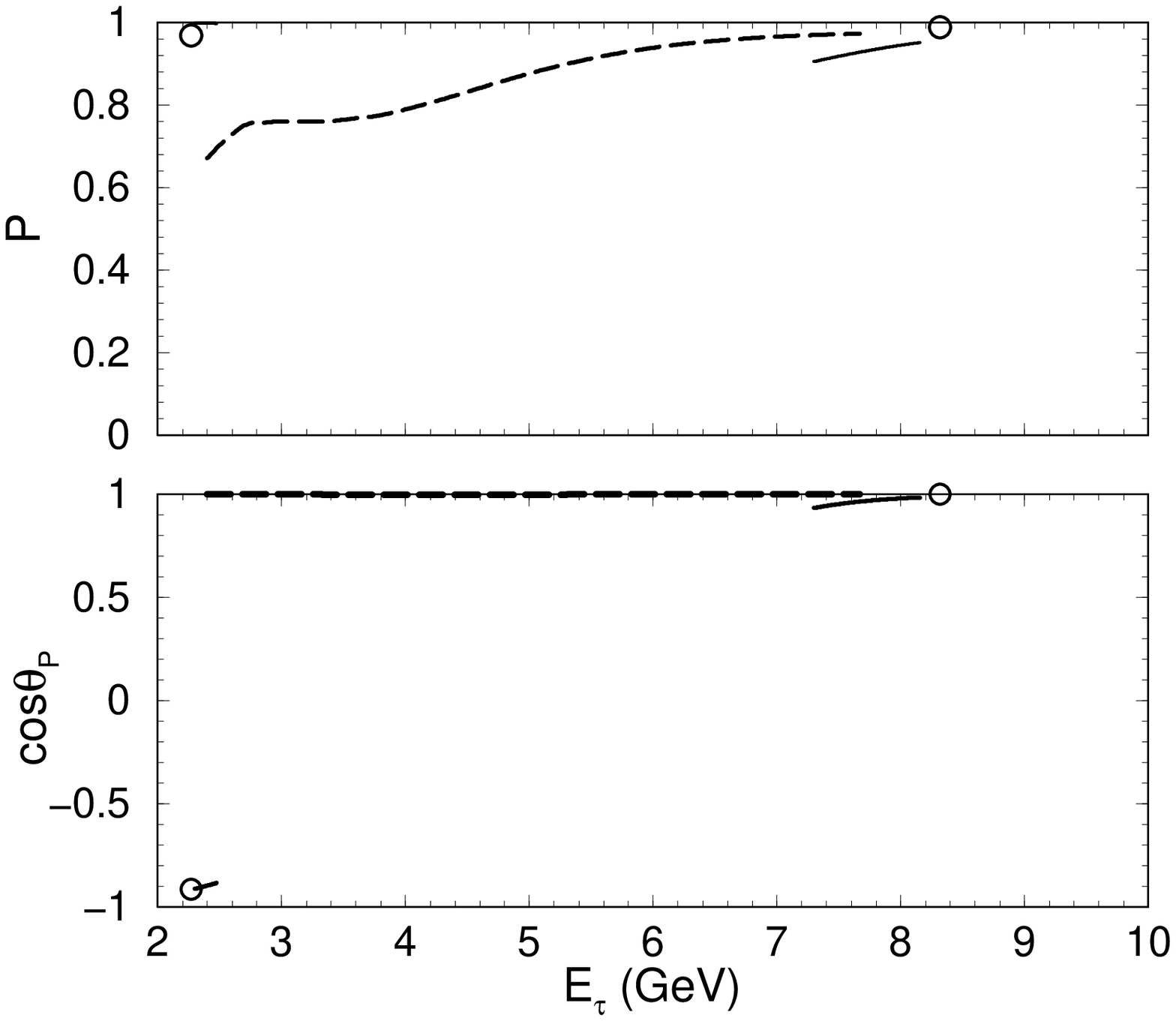,width=5cm}
\end{minipage}
\caption{The same as Fig.\ref{cross10}, but for the process 
$\bar{\nu}_{\tau}N \rightarrow \tau^{+}X$ at $E_{{\nu}}=10$GeV.
The $\tau^+$ spin in the $\tau^+$ rest 
frame is $\vec{s}={P\over 2}(\sin\theta_P,0,\cos\theta_P)$.
}\label{crossa10}
\end{figure}

Fig.\ref{crossa10} shows the $\bar{\nu}_{\tau}N \rightarrow \tau^{+}X$ 
case at $E_{\nu}=10$GeV. The predictions of the QE and RES processes are 
quite similar to those of the ${\nu}_{\tau}N \rightarrow \tau^{-}X$ 
process in Fig.\ref{cross10}, except that the  $\tau^+$ polarization is 
almost right-handed. In the DIS process, however,  
the polarization vector of $\tau^{+}$ is predicted to be quite different 
from the $\tau^{-}$ case in a non-trivial manner. 
This is because the $\bar{\nu}\bar{q}$ scattering contribution is not so 
small as compared to the $\bar{\nu}q$ scattering contribution. 
In case of the $\tau^+$ production process, 
the azimuthal angle of the $\tau^+$ polarization vector takes  
$\varphi_{P}=0$ at all energies, which gives $s_x={P\over 2}\sin\theta_P$
and $s_y=0$. Therefore the $\tau^+$ spin vector points away from  the 
initial neutrino beam axis, contrary to the $\tau^-$ spin case.  \\

\begin{figure}[H]
\begin{center}
\epsfig{figure=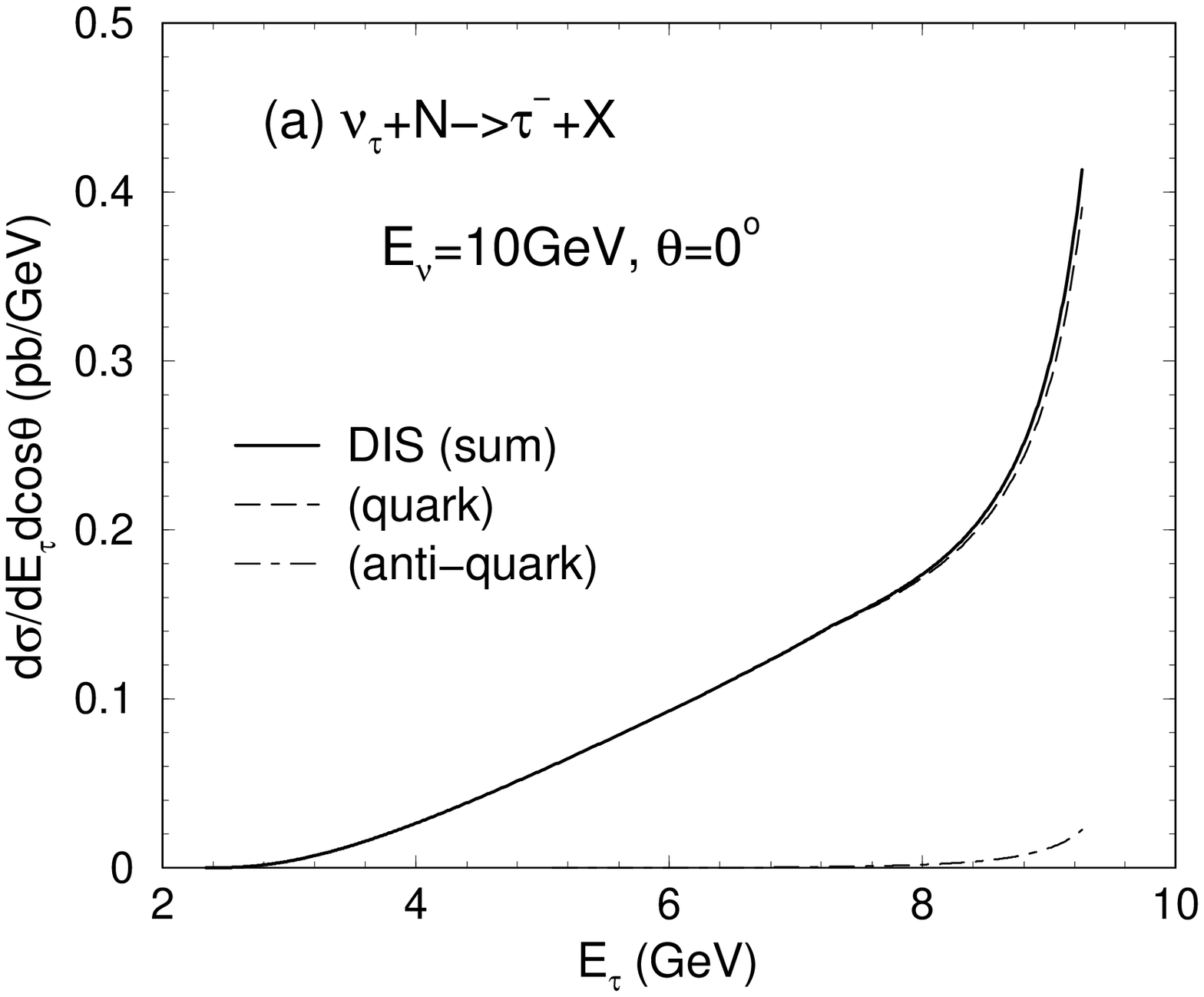,width=6.5cm}
\aki \epsfig{figure=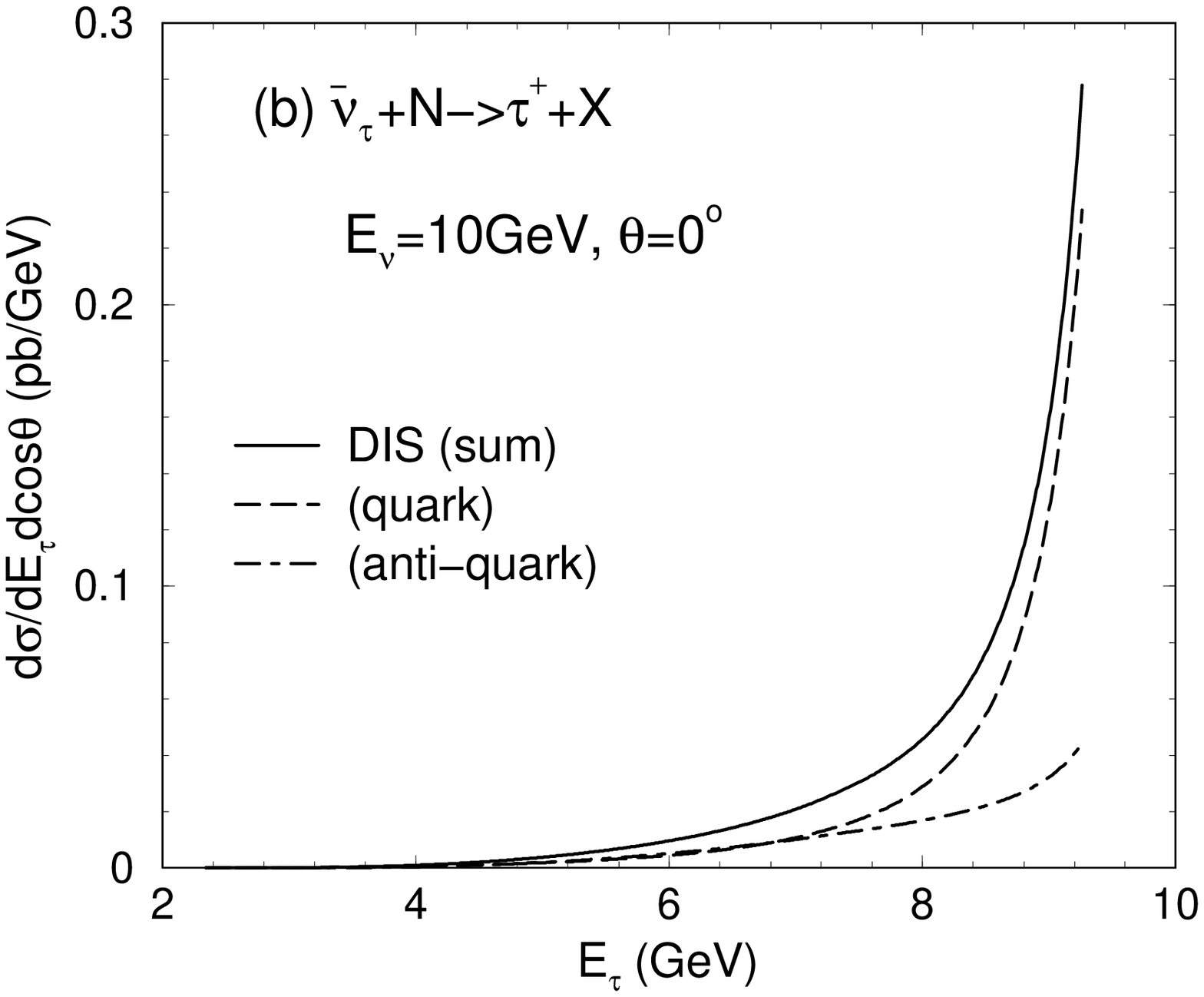,width=6.5cm}
\caption{
The differential cross section of the DIS process at $E_{\nu}=10$GeV 
and at $\theta=0^{\circ}$ in the laboratory frame for the processes 
$\nu_{\tau}N \rightarrow \tau^-X$ (a) and 
$\bar{\nu}_{\tau}N \rightarrow \tau^+X$ (b), where $N$ is an isoscalar 
nuclei. The dashed and dot-dashed lines represent, respectively, the 
contributions of the neutrino-quark and neutrino-antiquark 
scattering subprocesses. The solid lines show their sum.
}\label{quark}
\end{center}
\end{figure}

In order to understand the difference between the $\tau^+$ and $\tau^-$ 
spin polarization predictions in Fig.\ref{cross10} and Fig.\ref{crossa10}, 
we show in Fig.\ref{quark} the $E_{\tau}$ dependence of the differential 
cross section at $\theta=0^{\circ}$ for the 
${\nu}_{\tau}N \rightarrow \tau^{-}X$ process (a) and for the 
$\bar{\nu}_{\tau}N \rightarrow \tau^{+}X$ process (b). The contributions 
from the $\nu_{\tau}q$ or $\bar{\nu}_{\tau}q$ scattering process are 
shown by dashed lines, those from the  $\nu_{\tau}\bar{q}$ or 
$\bar{\nu}_{\tau}\bar{q}$ scattering process are shown by dash-dotted 
lines, and their sum by solid lines. It is clear that the 
${\nu}q$ scattering contribution dominates the 
${\nu}_{\tau}N \rightarrow \tau^{-}X$ process, whereas for the 
$\bar{\nu}_{\tau}N \rightarrow \tau^{+}X$ process, both  
$\bar{\nu} q$ or $\bar{\nu} \bar{q}$ scattering contribution are 
significant. \\

Let us consider the $\bar{\nu}q$ and 
$\bar{\nu}\bar{q}$ scattering in the parton collision CM frame. 
As for the $\bar{\nu}q$ scattering, the amplitude of right-handed 
$\tau^{+}$ production is proportional 
to $(1+\cos\hat{\theta})$ and that of left-handed $\tau^{+}$ 
production is proportional to $\sin\hat{\theta}$, where $\hat{\theta}$ 
is the scattering angle in the CM frame. On the other hand, 
for the $\bar{\nu}\bar{q}$ scattering the produced $\tau^{+}$ has 
fully right-handed polarization, and the angular distribution is flat 
in the CM frame. Next, let us consider the $\theta=0^{\circ}$ case 
in the laboratory frame, which correspond to $\hat{\theta}=0^{\circ}$ 
or $180^{\circ}$ in the CM frame. Because of the $(1+\cos\hat{\theta})^2$ 
and $\sin\hat{\theta}^2$ distributions in the CM frame, 
the $\tau^{+}$ from $\bar{\nu}q$ scattering has fully right-handed 
polarization and is produced 
only in the forward direction ($\hat{\theta}=0^{\circ}$) in the CM frame.
Hence, all $\tau^{+}$'s from $\bar{\nu}q$ scattering are right-handed 
along $\theta=0^{\circ}$ in the laboratory frame. 
On the other hand, the $\tau^{+}$'s from the $\bar{\nu}\bar{q}$ 
scattering are purely right-handed both at $\hat{\theta}=0^{\circ}$ 
and $180^{\circ}$ in the CM frame, and hence in the laboratory frame,  
the high energy $\tau^{+}$'s have right-handed polarization, 
and the low energy $\tau^{+}$'s have left-handed polarization.
By comparing the cross section of $\bar{\nu}q$ and $\bar{\nu}\bar{q}$ scattering 
in Fig.\ref{quark}(b), we find that high energy $\tau^+$'s are mostly 
produced by $\bar{\nu}q$ scattering, and hence they are almost right-handed. 
But as the energy decreases the contributions from 
$\bar{\nu}\bar{q}$ scattering increases and 
the degree of polarization becomes lower by cancellation.
In the QE and RES process of $\tau^{+}$ production, the mechanism is the
same as the $\tau^{-}$ production case but for the sign of  
the polarization. In the CM frame of $\bar{\nu}N$ scattering, 
$\tau^{+}$ has almost right-handed polarization for both QE and RES 
processes at all angles. Therefore in the laboratory frame,  
high energy $\tau^{+}$'s are right-handed, while low energy $\tau^{+}$'s are
left-handed because of the helicity flip by boost. \\

\begin{figure}[H]
\begin{minipage}{5cm}
\epsfig{figure=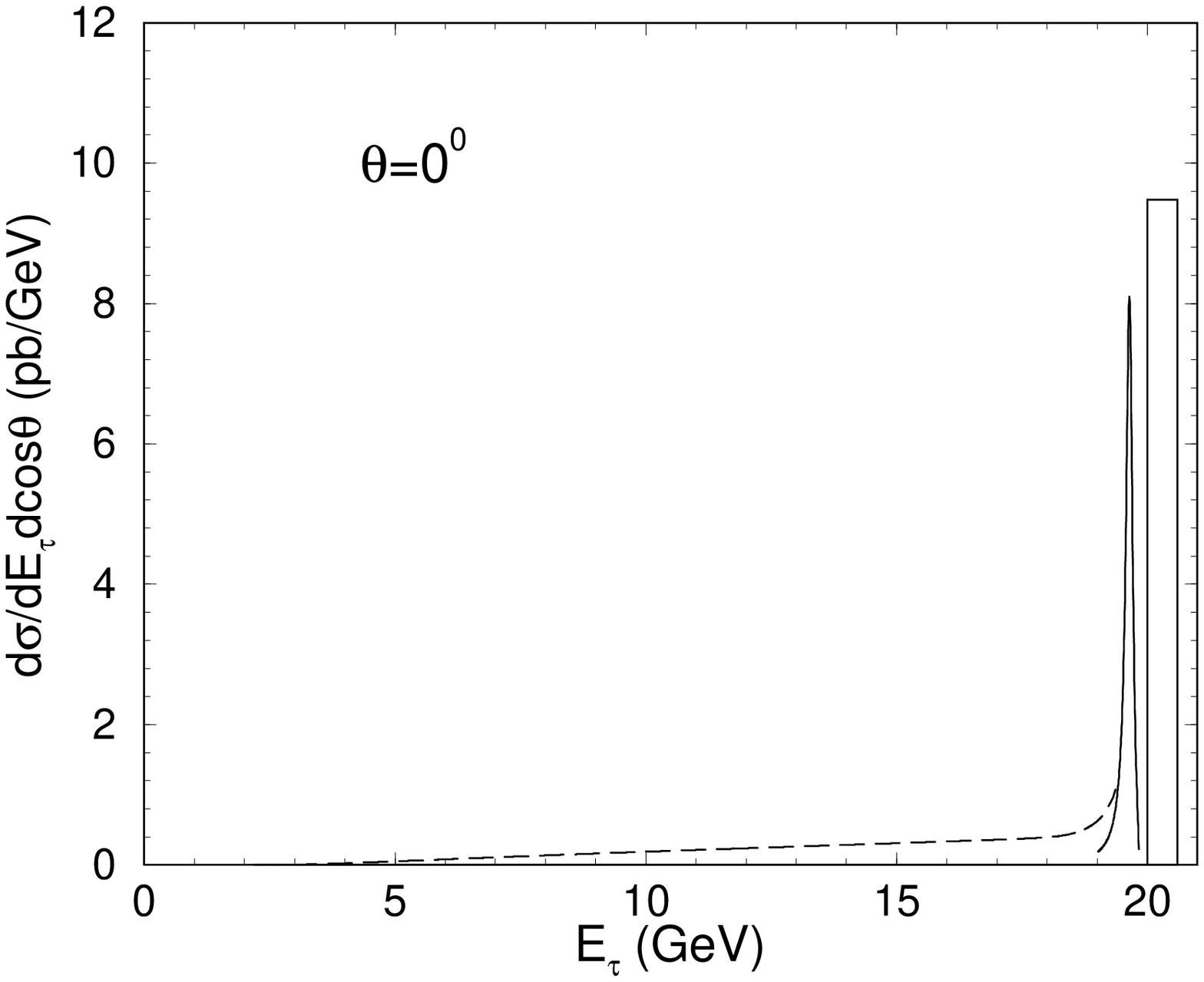,width=5cm}
\end{minipage}
\begin{minipage}{5cm}
\epsfig{figure=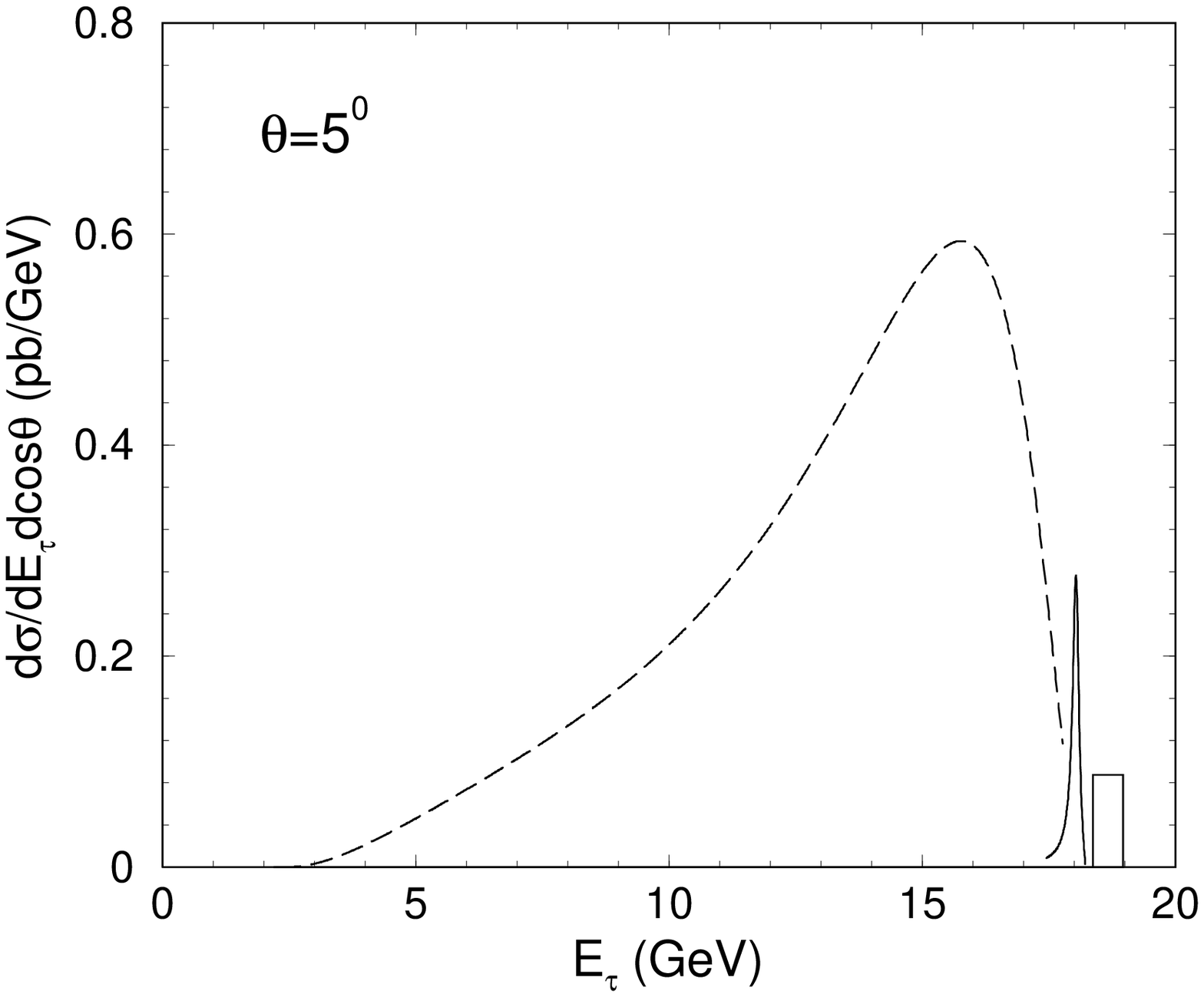,width=5cm}
\end{minipage}
\begin{minipage}{5cm}
\epsfig{figure=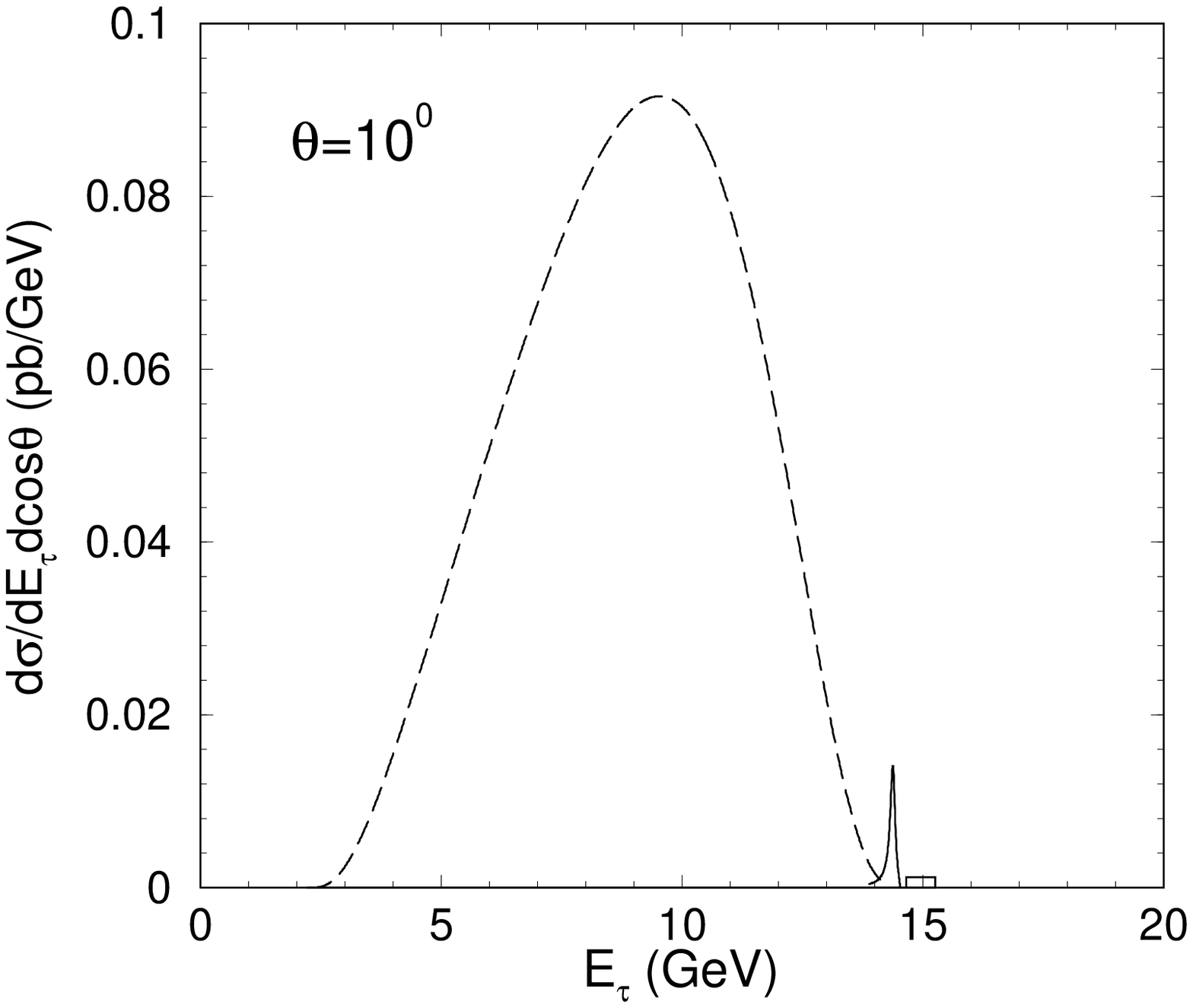,width=5cm}
\end{minipage}\\
\begin{minipage}{5cm}
\epsfig{figure=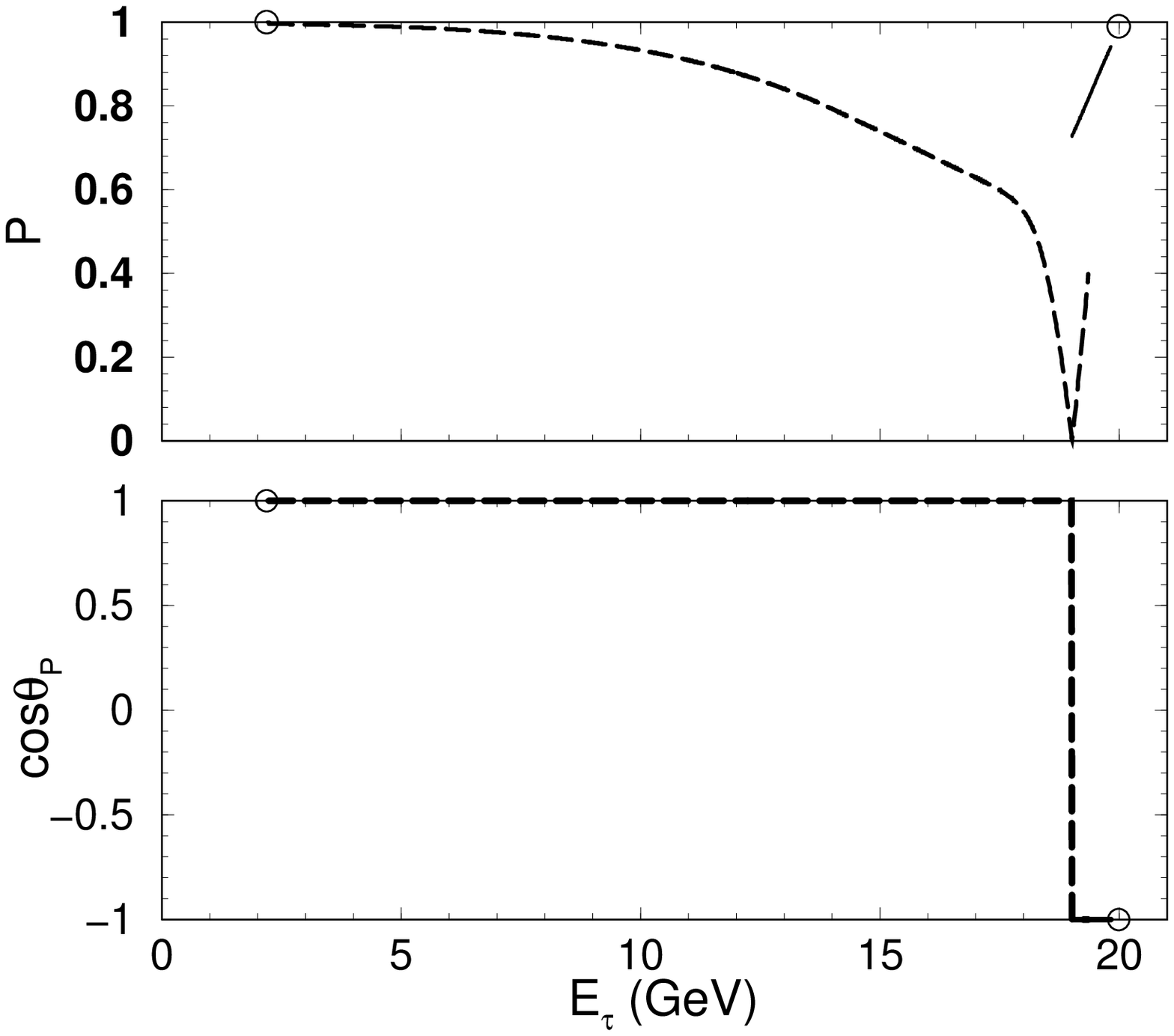,width=5cm}
\end{minipage}
\begin{minipage}{5cm}
\epsfig{figure=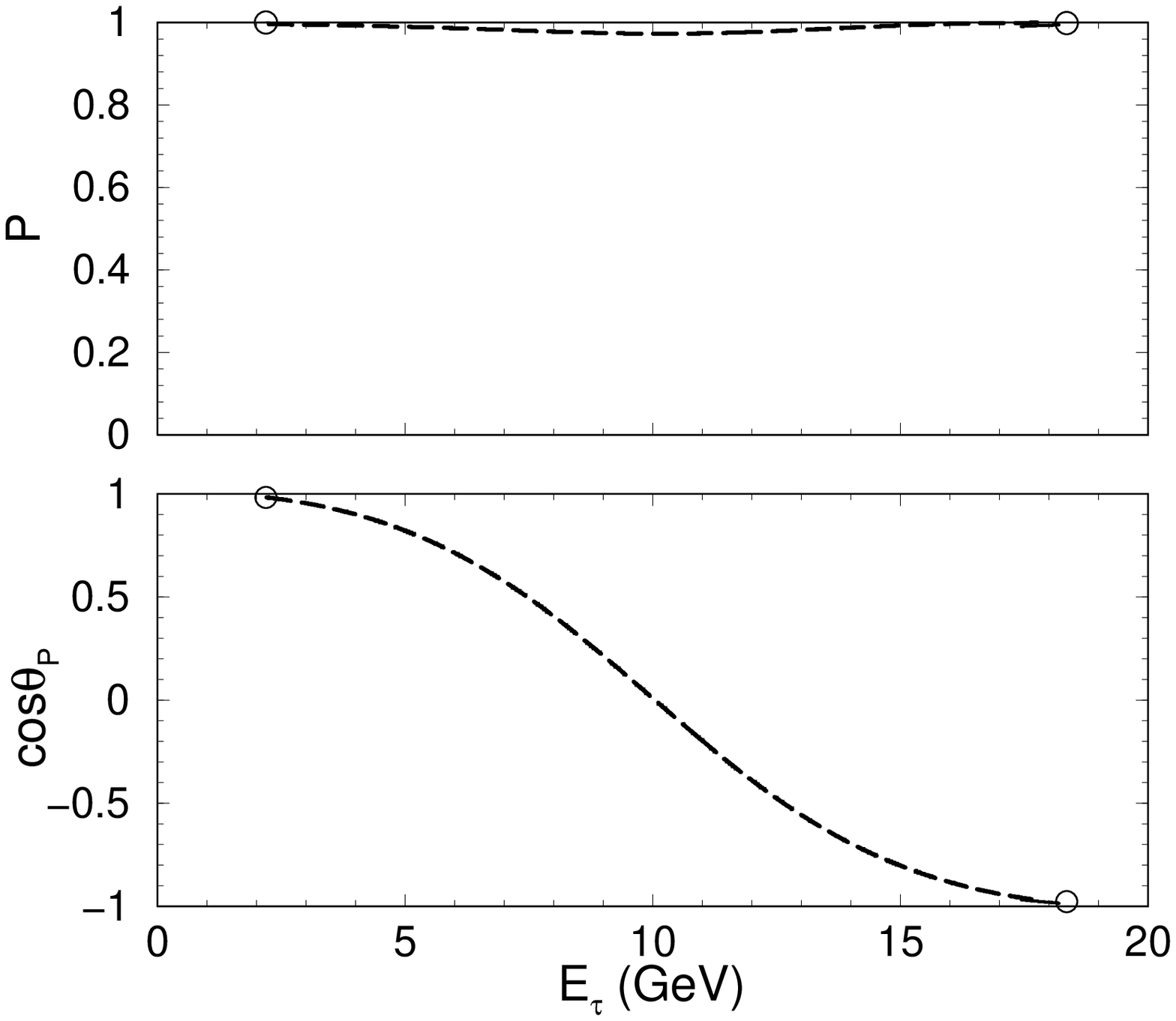,width=5cm}
\end{minipage}
\begin{minipage}{5cm}
\epsfig{figure=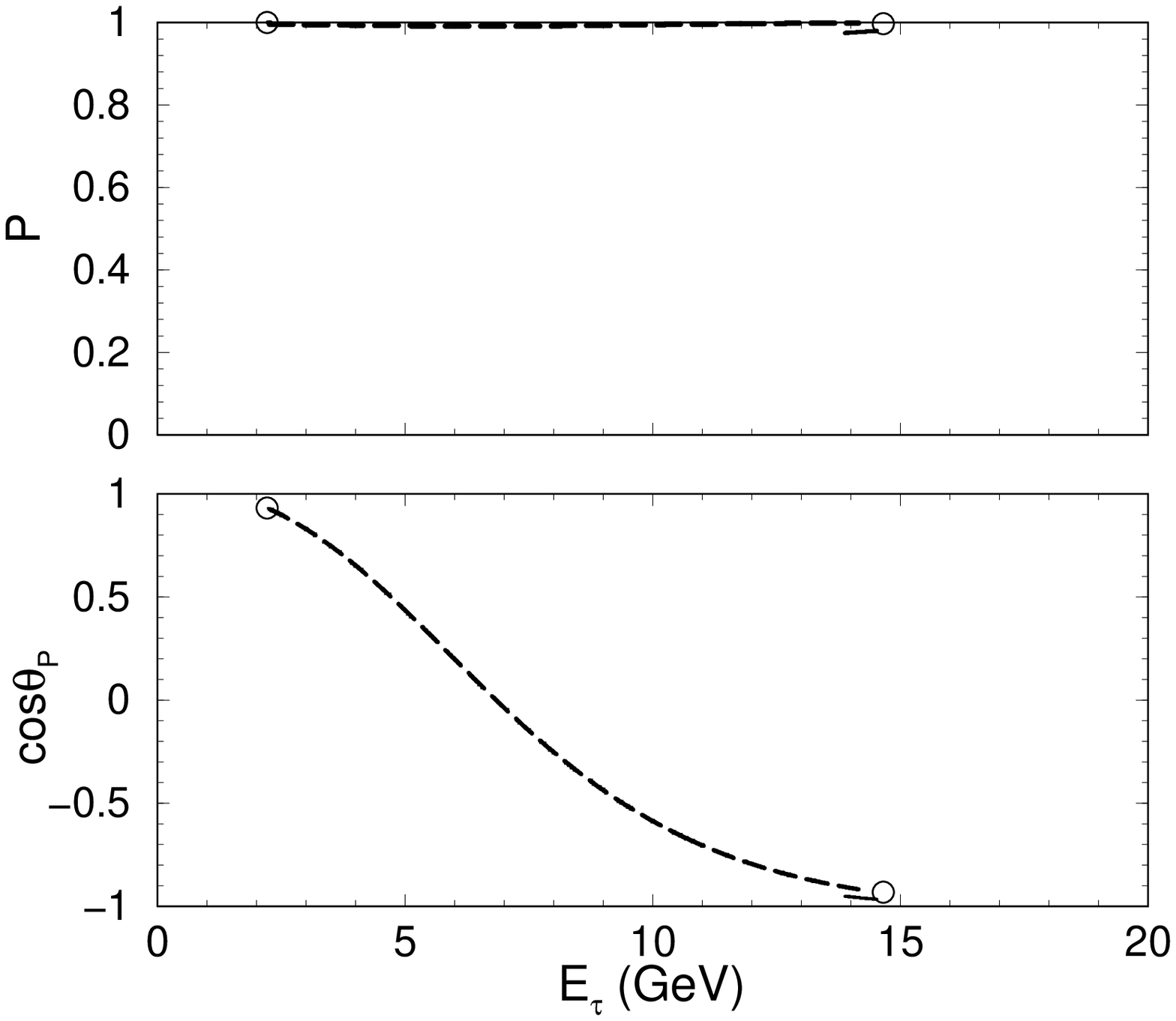,width=5cm}
\end{minipage}
\caption{
The same as Fig.\ref{cross10}, but for the process 
${\nu}_{\tau}N \rightarrow \tau^{-}X$ at $E_{{\nu}}=20$GeV.
}\label{cross20}
\end{figure}

\begin{figure}[H]
\begin{minipage}{5cm}
\epsfig{figure=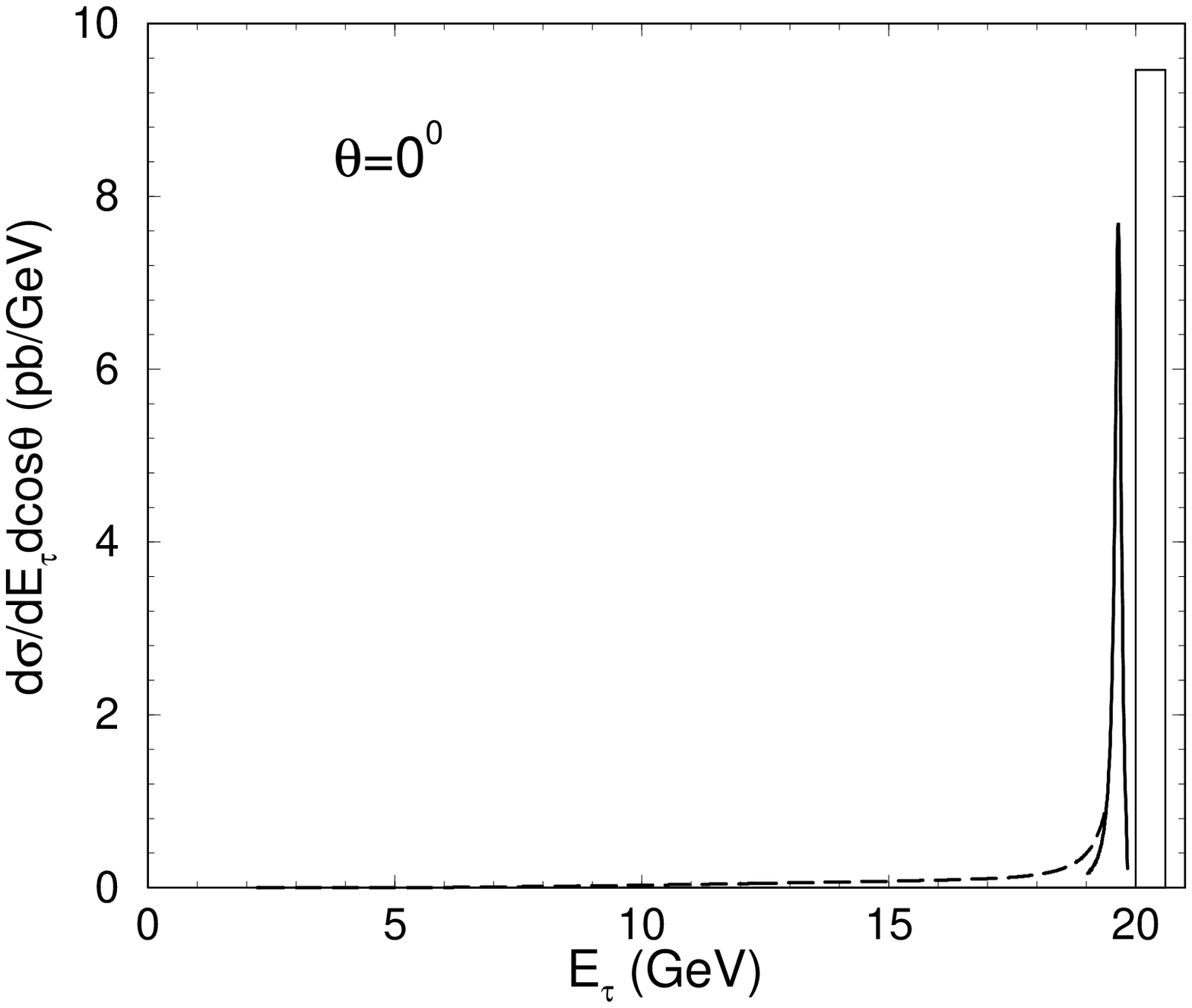,width=5cm}
\end{minipage}
\begin{minipage}{5cm}
\epsfig{figure=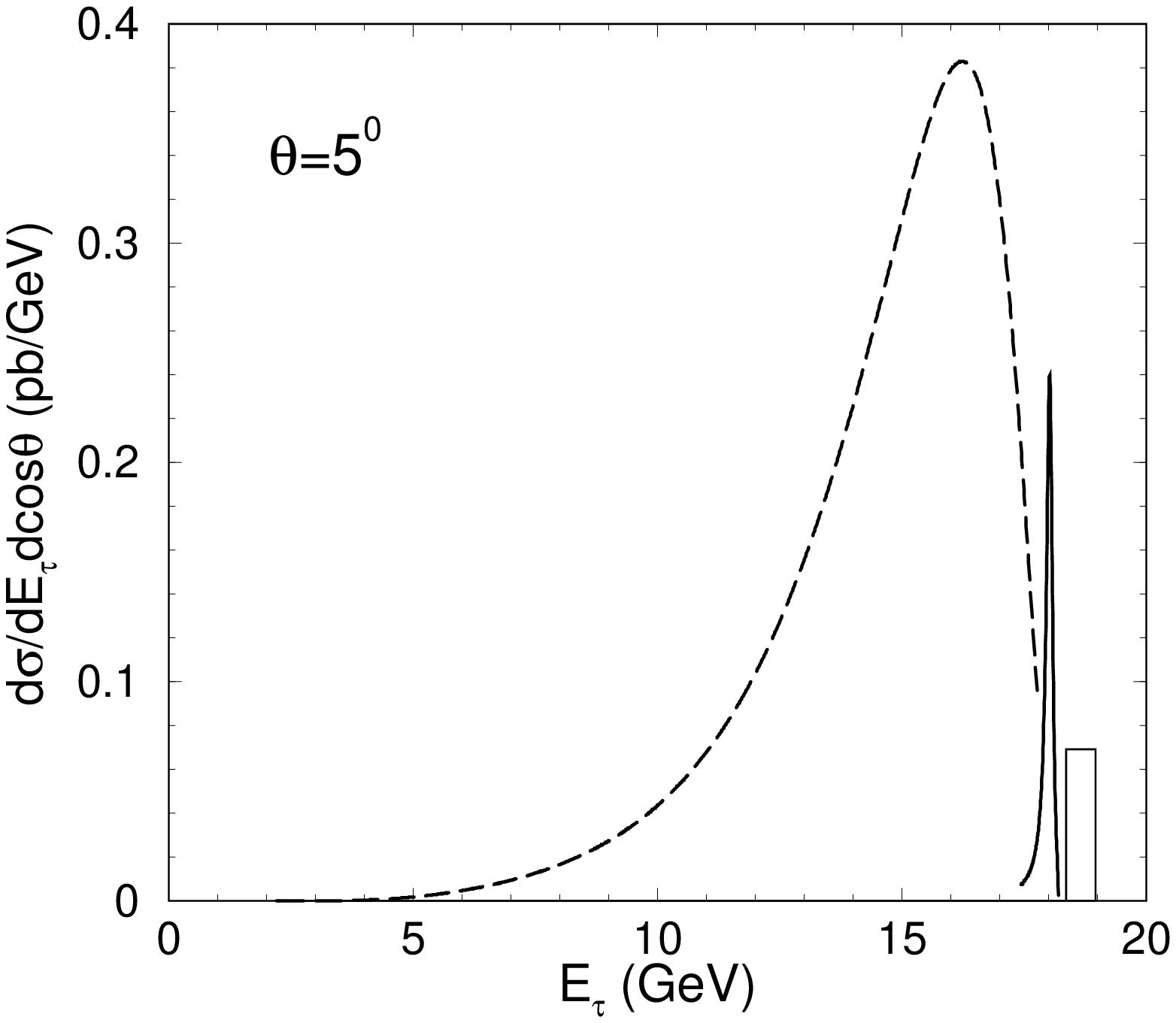,width=5cm}
\end{minipage}
\begin{minipage}{5cm}
\epsfig{figure=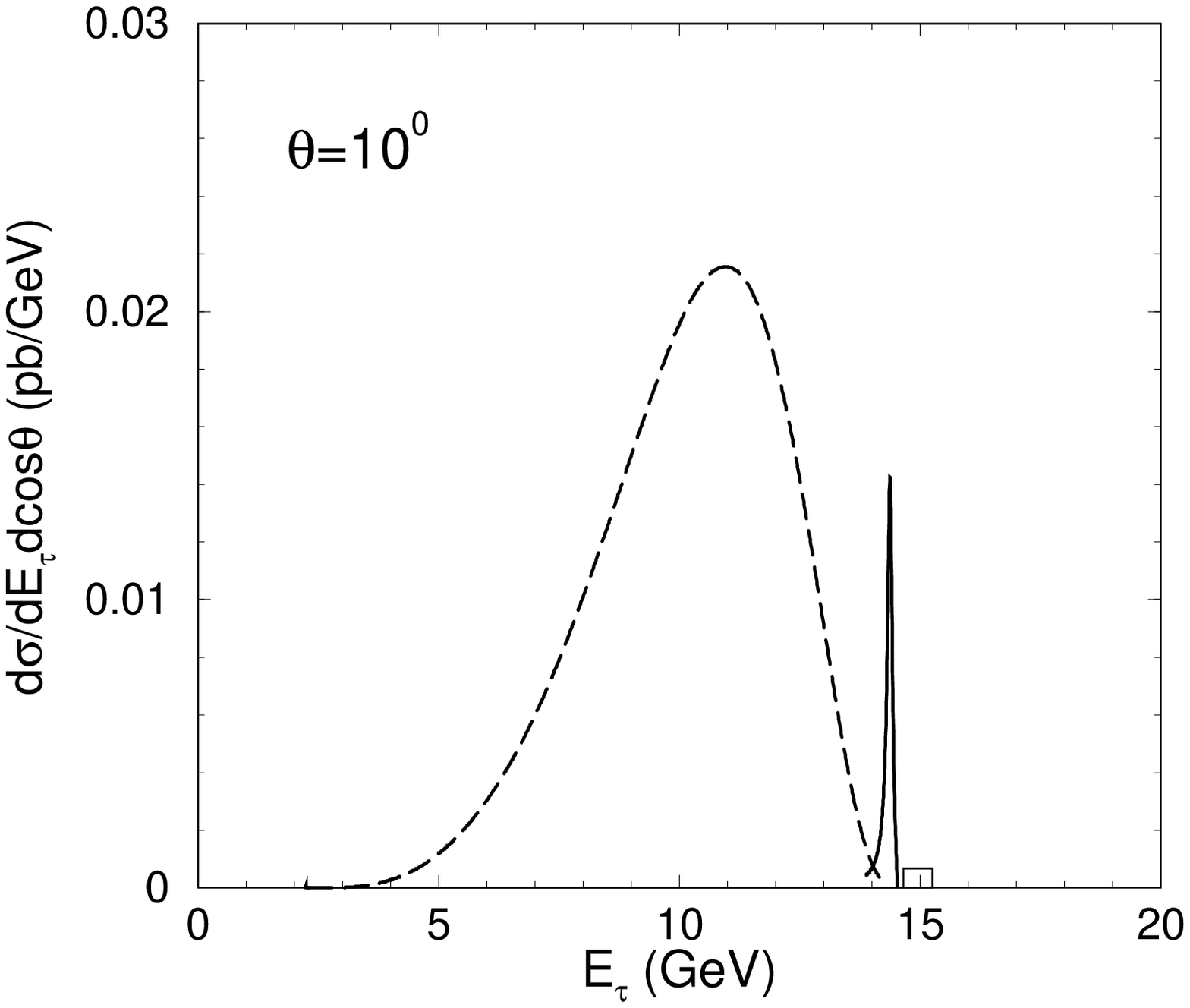,width=5cm}
\end{minipage}\\
\begin{minipage}{5cm}
\epsfig{figure=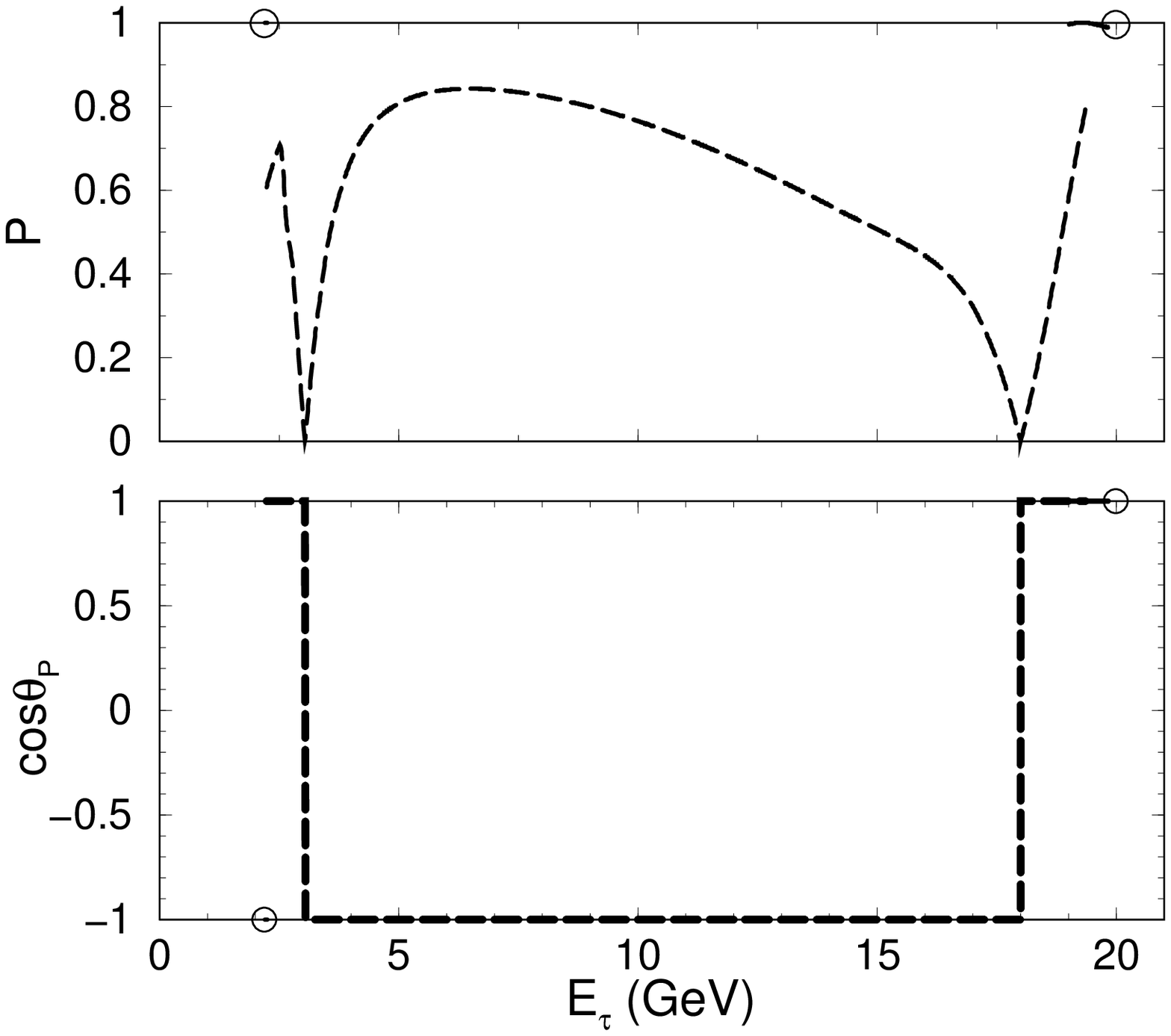,width=5cm}
\end{minipage}
\begin{minipage}{5cm}
\epsfig{figure=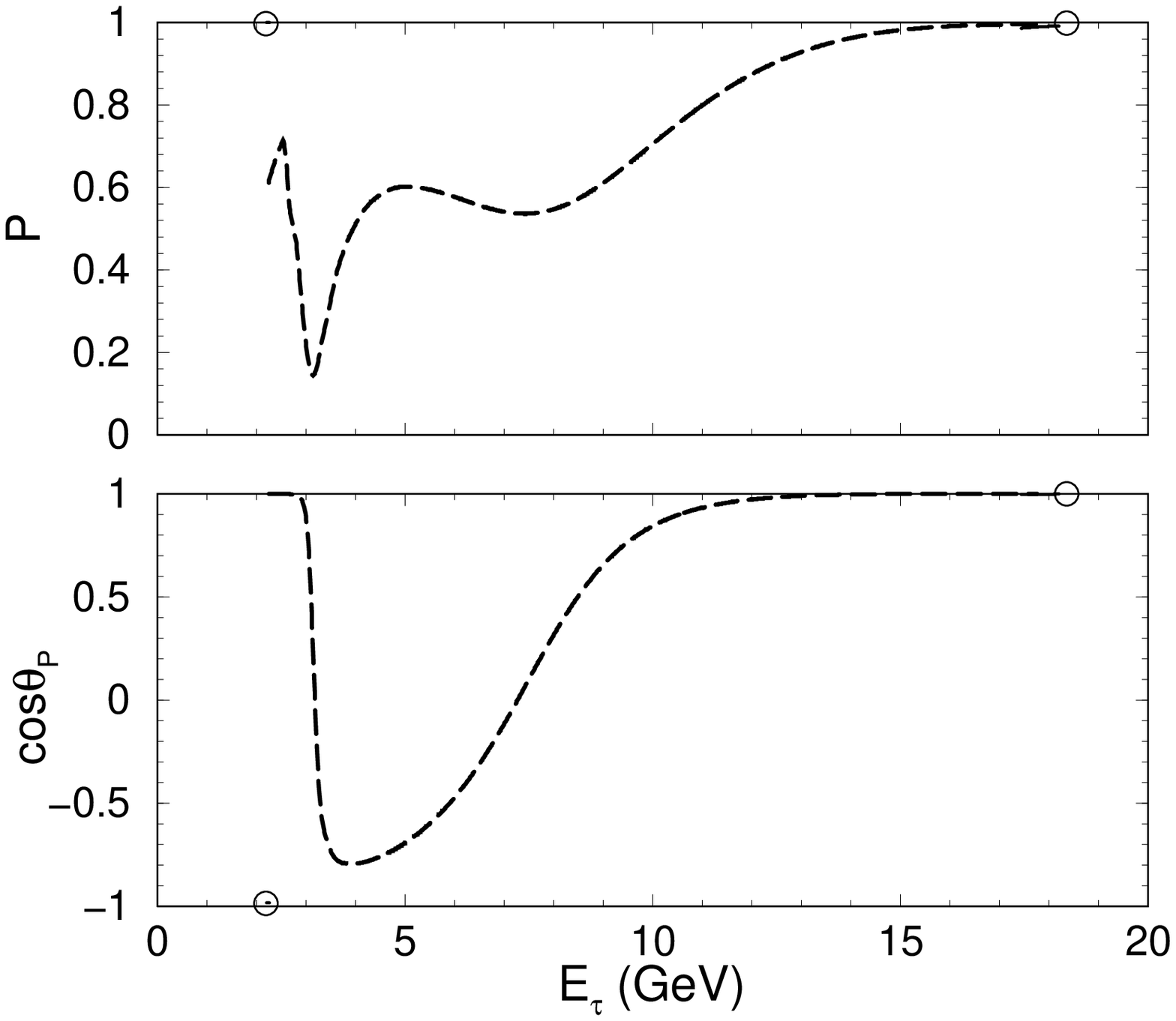,width=5cm}
\end{minipage}
\begin{minipage}{5cm}
\epsfig{figure=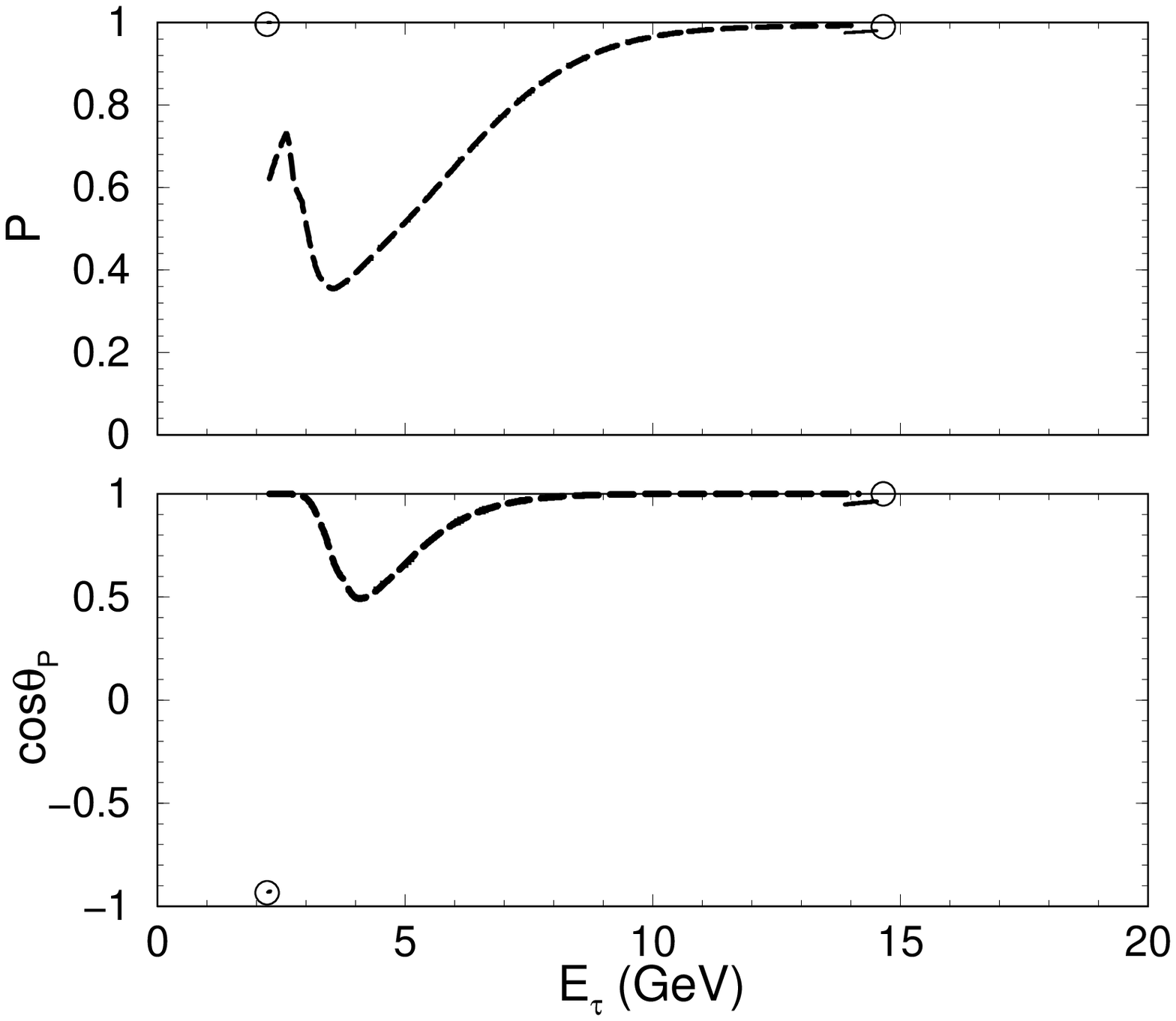,width=5cm}
\end{minipage}
\caption{The same as Fig.\ref{crossa10}, but for the process 
$\bar{\nu}_{\tau}N \rightarrow \tau^{+}X$ at $E_{{\nu}}=20$GeV.}
\label{crossa20}
\end{figure}

Let us now show our predictions at higher neutrino energies.
In Fig.\ref{cross20} and Fig.\ref{crossa20}, we show our predictions 
for $\nu_{\tau}N \rightarrow \tau^-X$ and 
$\bar{\nu}_{\tau}N \rightarrow \tau^+X$ processes, respectively, at 
$E_{\nu}=20$GeV. 
The energy dependence of the differential cross sections shows clearly the 
dominance of the DIS contribution at higher energies except at 
$\theta=0^{\circ}$. Fig.\ref{crossa20} shows that the QE and RES 
contributions are more significant in the $\bar{\nu}N$ scattering, as 
compared to the ${\nu}N$ scattering case shown in Fig.\ref{cross20}.
The relative importance of the QE or RES contribution to the
$\bar{\nu}N$ scattering persists at high energies as can be seen from  
Fig.\ref{total}. The degree of the $\tau^-$ polarization remains high in 
Fig.\ref{cross20}, except for the special angle of $\theta=0^{\circ}$, 
and its polarization direction are essentially understood by the boost 
effect, as for the $E_{\nu}=10$GeV case. In Fig.\ref{crossa20},
the degree of the $\tau^+$ polarization decreases at lower  $\tau^+$ 
energy in the laboratory frame for the 
$\bar{\nu}_{\tau}N \rightarrow \tau^+X$ process. 
This is understood as a result of the cancellation between the 
$\bar{{\nu}}q$ and $\bar{\nu}\bar{q}$ scattering contributions as in the 
$E_{\nu}=10$GeV case. \\

It is notable that the produced $\tau^-$ has almost 100\% 
polarization ($P\approx 1$) at all energies except at around 
$\theta=0^{\circ}$ while its polarization direction deviates from the 
pure left-handed direction ($\cos\theta_P=-1$) even at relatively high 
$\tau^-$ energies in the laboratory frame.
On the other hand, the $\tau^+$ polarization deviates from 100\% at 
relatively high $E_{\tau}$ while its direction is along the right-handed 
direction ($\cos\theta_P=+1$) down to half the maximum energy. 
Those qualitative difference between the $\tau^-$ polarization in the 
$\nu_{\tau}N \rightarrow \tau^-X$ process and the $\tau^+$ polarization 
in the $\bar{\nu}_{\tau}N \rightarrow \tau^+X$ process  is understood as 
a consequence of the significance of the antiquark contribution to the 
DIS process in $\bar{\nu}N$ scattering.

\section{Discussion and Conclusion}\label{discussion}
\hspace*{\parindent}
The information on the polarization of $\tau^\pm$ produced 
through the $\nu_{\tau}N$ and $\bar{\nu}_{\tau}N$ scattering is 
essential to identify 
the $\tau$ production signal since the decay particle distributions 
depend crucially on the $\tau$ spin. 
It is needed in long baseline neutrino oscillation experiments
which should verify the large $\nu_{\mu}\to \nu_{\tau}$ 
oscillation, and is also needed for the background estimation of 
$\nu_{\mu}\to\nu_{e}$ appearance experiments which should measure 
the small mixing angle of $\nu_{e}$-$\nu_{\mu}$ oscillation. \\

In this paper we studied in detail the spin polarization of 
$\tau^{\pm}$ produced in $\nu_{\tau}$ and $\bar{\nu}_{\tau}$ 
nucleon scattering via charged currents. 
Quasi-elastic scattering (QE), $\Delta$ resonance production (RES) and
deep inelastic scattering (DIS) processes have been studied.
The three subprocesses are distinguished by the hadronic invariant mass $W$. 
$W=M(=m_N)$ gives QE, $M+m_{\pi}<W<W_{\rm cut}$ gives RES and 
$W>W_{\rm cut}$ gives DIS. 
In this article, we set the kinematical boundary of RES and DIS process
at $W_{\rm cut}=1.4$GeV. \\

The spin density matrix of $\tau^\pm$ production 
has been defined and the $\tau^\pm$ spin polarization vector has been
defined and parametrized in the $\tau^\pm$ rest frame whose polar-axis 
is taken along the momentum direction of $\tau^\pm$ in the laboratory frame. 
The spin density matrix has been calculated for each subprocess by using the 
form factors for the QE and RES processes, and by using the parton 
distribution functions of Ref.\cite{mrst} for the DIS process.
We have shown the spin polarizations of $\tau^{\pm}$ as function of the 
$\tau^\pm$ energy and the scattering angle in the laboratory frame for 
$\nu_{\tau}N \rightarrow \tau^-X$ and 
$\bar{\nu}_{\tau}N \rightarrow \tau^+X$ processes
at $E_{\nu}=10$GeV and 20GeV.
We find that the produced $\tau^{\pm}$ have 
high degree of polarization, but their spin directions deviate significantly 
from the 
massless limit predictions at low and moderate $\tau$ energies.
Qualitative feature of the predictions have been 
understood by considering the helicity amplitudes 
in the CM frame of the scattering particles and the effects of Lorentz 
boost from the CM frame to the laboratory frame. \\

\begin{figure}[H]
\begin{center}
\epsfig{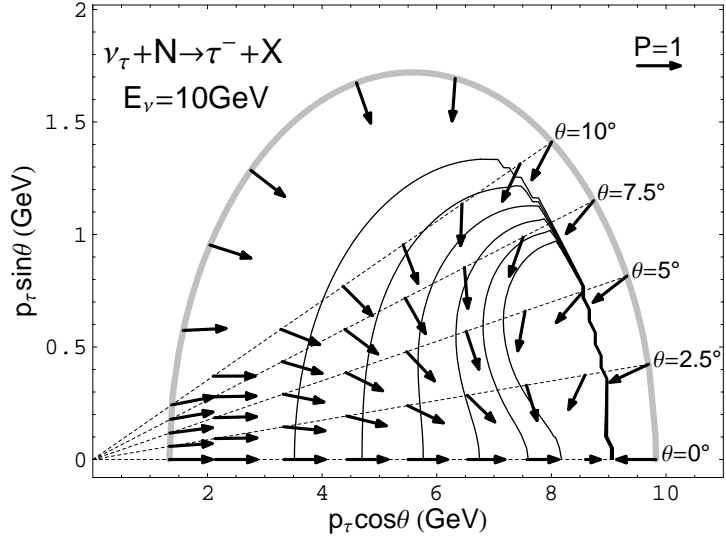}
\caption{The contour map of the DIS cross section in the plane of  
$p_{\tau}\cos\theta$ and $p_{\tau}\sin\theta$ for the  
$\nu_{\tau}N\to \tau^{-}X$ process at $E_{\nu}=10$GeV in the laboratory frame.
The kinematical boundary is shown by the thick grey curve. The QE process 
contributes along the boundary, and the RES process contributes just inside 
of the boundary. The $\tau^-$ polarization are shown by the arrows. 
The length of the arrows give the degree of
 polarization, and the direction of arrows give that of the $\tau^{-}$
spin in the $\tau^{-}$ rest frame. The size of the 100\% polarization ($P=1$) 
arrow is shown as a reference. The arrows are shown along the laboratory 
scattering angles, $\theta=0^{\circ}$, $2.5^{\circ}$, $5^{\circ}$, 
$7.5^{\circ}$, and $10^{\circ}$, as well as along the kinematical boundary.  
}\label{convec}
\end{center}
\end{figure}

\begin{figure}[H]
\begin{center}
\epsfig{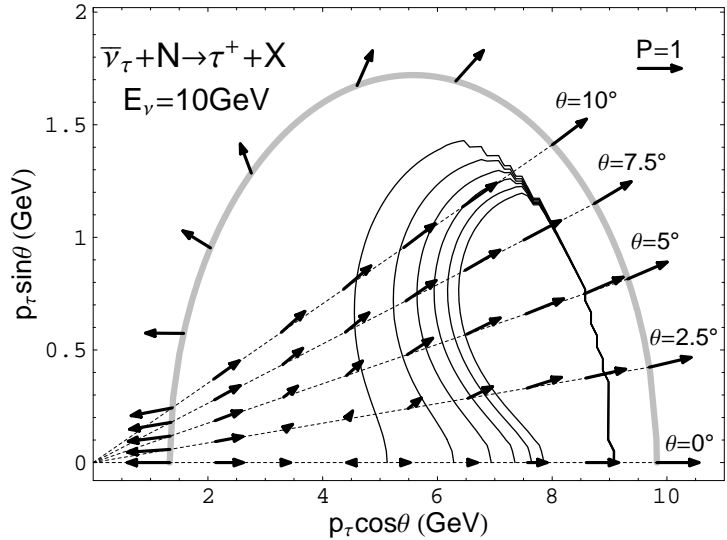}
\caption{The same as Fig.\ref{convec}, but for 
$\bar{\nu}_{\tau}N\to\tau^{+}X$ case.}\label{convecA}
\end{center}
\end{figure}

Finally, we summarize our findings in Fig.\ref{convec} and 
Fig.\ref{convecA}. 
In Fig.\ref{convec}, we show the polarization vector $\vec{s}$ of $\tau^{-}$
for the $\nu_{\tau}N \rightarrow \tau^-X$ process at $E_{\nu}=10$GeV 
on the $p_{\tau}\cos\theta$-$p_{\tau}\sin\theta$ plane, 
where $p_{\tau}$ and $\theta$ are the produced $\tau$ momentum 
and the scattering angle in the laboratory frame. 
The length of each arrow gives the degree of 
polarization ($0\leq P \leq 1$) at each phase-space point and its 
orientation gives the spin direction in the $\tau^-$ rest frame. 
The differential cross section is described as a contour map, 
where only the DIS cross section is plotted to avoid too much complexity.
The outer line gives the kinematical boundary, along 
which the QE process occurs.
Fig.\ref{convec} is a more visual version of the information given in 
Fig.\ref{cross10}.
Fig.\ref{convecA} gives the $\tau^{+}$ polarization 
for the $\bar{\nu}_{\tau}N \rightarrow \tau^+X$ process at $E_{\nu}=10$GeV, 
compiling the cross section and the polarization information of 
Fig.\ref{crossa10}.\\

Before closing our discussion, we point out some uncertainties
in our calculation.
One is the uncertainty at small $Q^2$ 
region ($Q^2 < 1$GeV$^2$) in our DIS calculation.
In this paper, we used an extrapolation of 
the parton model calculation in this region by freezing the PDF's below
their validity region.
Because the parton model must break down in this region, and because
our estimation of the cross section 
in this region is not small, a more careful treatment, e.g. by using the 
structure function data is needed.
Another is the uncertainty in the pseudo-scalar form factors, 
$F_p(q^2)$ for the QE and $C^A_6(q^2)$ for the RES processes, 
which are not known enough so far. 
Because of the large $\tau$ mass and because of the spin-flip nature of 
those form factors, they can affect the predictions of the $\tau^{\pm}$ 
polarization significantly.  QCD higher-order corrections should affect 
the $\tau^{\pm}$ polarization in the DIS region. We plan to study those 
uncertainties elsewhere. \\

We hope that this work will be useful in detecting the $\tau$ appearance 
signal in long baseline neutrino oscillation experiments, and 
that it will also be useful in understanding the 
$\tau^{\pm} \to l^{\pm}\nu\bar{\nu}$ background for the 
$\nu_{\mu} \to \nu_e$ appearance experiments.

\vspace*{-1cm}

\begin{center}
{\bf \Large Acknowledgments}\\
\end{center}
We are grateful to E.A.Paschos for useful comments and discussions.
K.M. and H.Y. thank KEK theory group for the hospitality, where parts of 
this work were performed. 
K.M. would like to thank T.Morii and S.Oyama for discussions.
 H.Y. would like to thank M.Hirata, J.Kodaira for discussions,
and RIKEN BNL Research Center for the hospitality
 where this work was finalized.


\end{document}